\documentclass[submission,a4paper,Phys]{SciPost}

\pdfoutput=1
\usepackage{amsmath,amssymb,mathtools,xspace}
\usepackage{booktabs,multirow,graphicx,tabularx,slashed}
\usepackage{hyperref}
\usepackage{color,xcolor}
\usepackage[normalem]{ulem}
\usepackage{enumitem}
\usepackage{feynmp}
\usepackage{braket}
\usepackage{tikz}
\usetikzlibrary{shapes.geometric, arrows}

\makeatletter
\@ifundefined{pdfoutput}{}{\DeclareGraphicsRule{*}{mps}{*}{}}
\makeatother

\definecolor{Gcolor}{HTML}{3b528b}
\definecolor{Dcolor}{HTML}{e41a1c}

\tikzstyle{generator} = [rectangle, rounded corners, minimum width=3cm, minimum height=1cm,text centered, draw=Gcolor]
\tikzstyle{discriminator} = [rectangle, rounded corners, minimum width=3cm, minimum height=1cm,text centered, draw=Dcolor]
\tikzstyle{io} = [circle, trapezium left angle=70, trapezium right angle=110, minimum width=1cm, minimum height=1cm, text centered, draw=black]

\tikzstyle{process} = [rectangle, minimum width=1cm, minimum height=1cm, text centered, draw=black]
\tikzstyle{decision} = [rectangle, minimum width=1cm, minimum height=1cm, text centered, draw=black]

\tikzstyle{arrow} = [thick,->,>=stealth]
\usepackage{xcolor}

\newcommand{\Langle}{\big\langle}
\newcommand{\Rangle}{\big\rangle}

\newcommand\one{\leavevmode\hbox{\small1\normalsize\kern-.33em1}}

\newcommand{\mat}{\mathcal{M}}

\def\slashchar#1{\setbox0=\hbox{$#1$}           
   \dimen0=\wd0                                 
   \setbox1=\hbox{/} \dimen1=\wd1               
   \ifdim\dimen0>\dimen1                        
      \rlap{\hbox to \dimen0{\hfil/\hfil}}      
      #1                                        
   \else                                        
      \rlap{\hbox to \dimen1{\hfil$#1$\hfil}}   
      /                                         
   \fi}

\renewcommand{\d}{{\text{d}}}

\newcommand{\vegas}{\texttt{Vegas}\xspace}
\newcommand{\madgraph}{\texttt{MG5aMCNLO}\xspace}

\begin{document}
\begin{fmffile}{feynman}

\begin{center}{\Large \textbf{
How to GAN LHC Events
}}\end{center}

\begin{center}
Anja Butter\textsuperscript{1},
Tilman Plehn\textsuperscript{1}, and
Ramon Winterhalder\textsuperscript{1}
\end{center}

\begin{center}
{\bf 1} Institut f\"ur Theoretische Physik, Universit\"at Heidelberg, Germany
winterhalder@thphys.uni-heidelberg.de
\end{center}

\begin{center}
\today
\end{center}


\section*{Abstract}
{\bf Event generation for the LHC can be supplemented by generative
  adversarial networks, which generate physical events and avoid highly
  inefficient event unweighting.  For top pair production we show how
  such a network describes intermediate on-shell particles, phase
  space boundaries, and tails of distributions. In particular, we
  introduce the maximum mean discrepancy to resolve sharp local
  features. It can be extended in
  a straightforward manner to include for instance off-shell
  contributions, higher orders, or approximate detector effects.}

\vspace{10pt}
\noindent\rule{\textwidth}{1pt}
\tableofcontents\thispagestyle{fancy}
\noindent\rule{\textwidth}{1pt}
\vspace{10pt}

\newpage
\section{Introduction}
\label{sec:intro}

First-principle simulations are a key ingredient to the ongoing
success of the LHC, and they are crucial for further developing it
into a precision experiment testing the structure of the Standard
Model and its quantum field theory underpinnings. Such simulations of
the hard scattering process, QCD activity, hadronization, and detector
effects are universally based on Monte Carlo methods. These methods
come with structural challenges, for example related to an efficient
coverage of the high-dimensional phase space, event unweighting, or
complex and hence slow detector simulations. Some of these problems
might be alleviated when we add a new direction, like machine learning
techniques, to our tool box. While we should not expect them to
magically solve all problems, we have seen that modern machine
learning can trigger significant progress in LHC physics. The reason
for our optimism related to event generation are generative
adversarial networks or GANs~\cite{goodfellow}, which have shown
impressive performance in tasks like the generation of images, videos
or music.

From the experimental side the detector simulation is the most
time-consuming aspect of LHC simulations, and promising attempts exist
for describing the behavior of the calorimeter with the help of
generative
networks~\cite{calogan1,calogan2,fast_accurate,aachen_wgan,ATLASShowerGAN,ATLASsimGAN}. On
the theory side, we know that the parton shower can be described by a
neural network~\cite{shower,locationGAN,monkshower,juniprshower}. It
has been shown that neural networks can help with phase space
integration~\cite{maxim,bendavid} and with LHC event
simulations~\cite{dutch,gan_datasets,DijetGAN}. One open question is
why the GAN setup of Ref.~\cite{dutch} does not properly work and is
replaced by a variational autoencoder with a density information
buffer. Another challenge is how to replace the ad-hoc $Z$-constraint
in the loss function of Ref.~\cite{gan_datasets} by a generalizable
approach to on-shell resonances. This problem of intermediate
resonances is altogether avoided in Ref.~\cite{DijetGAN}. It remains to
be shown how GANs can actually describe realistic multi-particle
matrix elements over a high-dimensional phase space in a flexible and
generalizable manner.

In this paper we show how we can efficiently GAN\footnote{From
  `to GAN', in close analogy to the verbs taylor, google, and
  sommerfeld.} the simulation of the $2 \to 6$ particle production
process
\begin{align}
pp \to t \bar{t} \to (b q \bar{q}') \; (\bar{b} \bar{q} q') 
\end{align}
describing all intermediate on-shell states with Breit-Wigner
propagators and typical width-to-mass ratios of few per-cent. We will
focus on a reliable coverage of the full phase space, from simple
momentum distributions to resonance peaks, strongly suppressed
tails, and phase space boundaries.

Given this new piece of the event simulation puzzle through fast
neural networks it should in principle be possible to add parton
showers, possibly including hadronization, and detector effects to a
full machine learning description of LHC events. Including
higher-order corrections is obviously possible and should lead to ever
higher gains in computing time, assuming higher-orders are included in the training data. The interesting question then becomes
where established methods might benefit from the fast and efficient
machine learning input. Alternatively, we can replace the Monte Carlo
event input and instead generate reconstructed LHC events and use them to
enhance analyses or to study features of the hard process. Obviously,
the GAN approach also allows us to combine information from actual
data with first-principles simulations in a completely flexible
manner.

Our paper consists of two parts.  In Sec.~\ref{sec:ps} we start by
reviewing some of the features of phase space sampling with Monte
Carlo methods and introducing GANs serving the same purpose. We then add 
the MMD and describe how its been used to describe intermediate resonances. In
Sec.~\ref{sec:ps_gan} we apply the combined GAN-MMD network to top
pair production with subsequent decays and show that it describes the
full phase space behavior, including intermediate on-shell particles.

\section{Phase space generation}
\label{sec:ps}

\begin{figure}[b!]
\begin{center}
\input{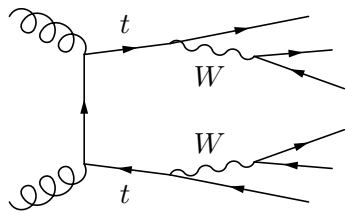}
\end{center}
\caption{Sample Feynman diagram contributing to top pair production,
  with intermediate on-shell particles labelled.}
\label{fig:feyn_intro}
\end{figure}

As a benchmark model throughout this paper we rely on top pair
production with an intermediate decay of two $W$-bosons
\begin{align}
pp
\to t\bar{t} 
\to (b W^-) \; (\bar{b} W^+) 
\to (b f_1 \bar{f}_1') \; (\bar{b} f_2 \bar{f}_2') \; ,
\end{align}
illustrated in Fig.~\ref{fig:feyn_intro}. If we assume that the masses
of all final-state particles are known, as this can be extracted from the measurement, this leaves us with 18 degrees
of freedom, which energy-momentum conservation reduces to a
14-dimensional phase space. In addition, our LHC simulation has to
account for the 2-dimensional integration over the parton momentum
fractions.

In this section we will briefly review how standard methods describe
such a phase space, including the sharp features of the intermediate
on-shell top quarks and $W$-boson. The relevant area in phase space is
determined by the small physical particle widths and extends through a
linearly dropping Breit-Wigner distribution, where it eventually needs to
include off-shell effects. We will then show how a generative
adversarial network can be constructed such that it can efficiently
handle these features as well.

\subsection{Standard Monte Carlos}
\label{sec:standard}

For the hard partonic process we denote the incoming parton momenta as
$p_{a,b}$ and the outgoing fermion momenta as $p_f$.  The partonic
cross section and the 14-dimensional phase-space integration for six
external particles can be parametrized as
\begin{align}
\int\d \sigma&=\int\d \Phi_{2\to6}\,\dfrac{\vert\mat (p_a,p_b;p_1,\dots,p_6)\vert^2}{2\hat{s}} \notag \\
\text{with} \quad
\d \Phi_{2\to6}&=\left.(2\pi)^4\delta^{(4)}(p_a+p_b-p_1- \dots-p_6)\prod_{f=1}^{6} \frac{\d ^3p_f}{(2\pi)^3}\frac{1}{2p_f^0}\right\vert_{p_f^0=\sqrt{\vec p_f^{\,2}+m_f^2}} \; .
\label{eq:def_ps}
\end{align}
To cope with the high dimensionality of the integral we adopt
advanced Monte Carlo techniques. The
integral of a function $f(x)$ over a volume $V$ in $\mathbb{R}^d$
\begin{align}
I=\int_V\d ^dx\,f(x) 
\end{align}
can be approximated with the help of $N$ random numbers $x_i$
distributed according to a normalized density function
$\rho(x)$
\begin{align}
\int_V\d ^dx\,\rho(x)=1 \; ,
\end{align}
such that
\begin{align}
I \approx S_N =\frac{1}{N}\sum_{i=1}^N \frac{f(x_i)}{\rho(x_i)}  \; .
\end{align}
For sufficiently large $N$ the variance of this integral scales like
\begin{align}
\sigma^2 \approx 
\frac{1}{N-1}\left(\frac{1}{N}\sum_{i=1}^{N} \frac{f(x_i)^2}{\rho(x_i)^2} -S^2_N\right) \; ,
\end{align}
which means that it can be minimized by an appropriate choice of
$\rho(x)$. This requires $\rho(x)$ to be large in regions where the
integrand is large, for instance
\begin{align}
\rho(x)
=\dfrac{\vert f(x)\vert}{\int_V\d^dx\,f(x)} \; .
\end{align} 
This method of choosing an adequate density is called importance
sampling. There are several implementations available, one of the most
frequently used is \vegas~\cite{vegas1,vegas2}.  

A major challenge in particle physics applications is that
multi-particle amplitudes in the presence of kinematic cuts typically have
dramatic features. Our phase space sampling not only
has to identify the regions of phase space with the leading
contribution to the integral, but also map its features
with high precision. For instance, the process
illustrated in Fig.~\ref{fig:feyn_intro} includes narrow intermediate
on-shell particles. Around a mass peak with $\Gamma \ll m$ they lead
to a sharp Breit-Wigner shape of the transition amplitude.  A standard
way of improving the integration is to identify the invariant mass
variable $s$ where the resonance occurs and switch variables to 
\begin{align}
\int \d s \frac{F(s)}{(s-m^2)^2 + m^2 \Gamma^2}
= \frac{1}{m \Gamma} \int \d z \; F(s) 
\quad \text{with} \quad 
z = \arctan \frac{s-m^2}{m\Gamma} \; .
\label{eq:ps_mapping}
\end{align}
This example illustrates how phase space mappings,
given some knowledge of the structure of the integrand, allow us to
evaluate high-multiplicity scattering processes.

Finally, in LHC applications we are typically not interested in an
integral like the one shown in Eq.\eqref{eq:def_ps}. Instead, we want
to simulate phase space configurations or events with a probability
distribution corresponding to a given hard process, shower
configuration, or detector smearing. This means we have to transfer
the information included in the weights at a given phase space point to a
phase space density of events with uniform weight. The corresponding
unweighting procedure computes the ratio of a given event weight to
the maximum event weights, probes this ratio with a random number, and
in turn decides if a phase space point or event remains in the sample,
now with weight one. This procedure is highly inefficient.

Summarizing, the challenge for a machine learning approach to phase
space sampling is: mimic importance sampling, guarantee a precise
mapping of narrow patterns, and avoid the limited unweighting
efficiency.

\subsection{Generative adversarial network}
\label{sec:gan}

The defining structural elements of generative adversarial networks or
GANs are two competing neural networks, where the generator network $G$
tries to mimic the data while the discriminator network $D$ is trained
to distinguish between generated and real data. The two networks play
against each other, dynamically improving the generator by searching
for parameter regions where the generator fails and adjusting its
parameters there. 

To start with, both networks are initialized with random values so
that the generator network induces a underlying random distribution
$P_G(x)$ of an event or phase space configuration $x$, typically
organized with the same dimensionality as the (phase) space we want to
generate. Now the discriminator network compares two distributions,
the true distribution $P_T(x)$ and the generated distribution
$P_G(x)$.  From each of the two distributions we provide batches of
phase space configurations $\{ x_T \}$ and $\{ x_G\}$ sampled from
$P_T$ or $P_G$, respectively. Here the sets $\{ x_{T,G} \}$ are
batches of events sampled from the training or generated data.

\begin{figure}[t]
\centering
\includegraphics[width=0.45\textwidth]{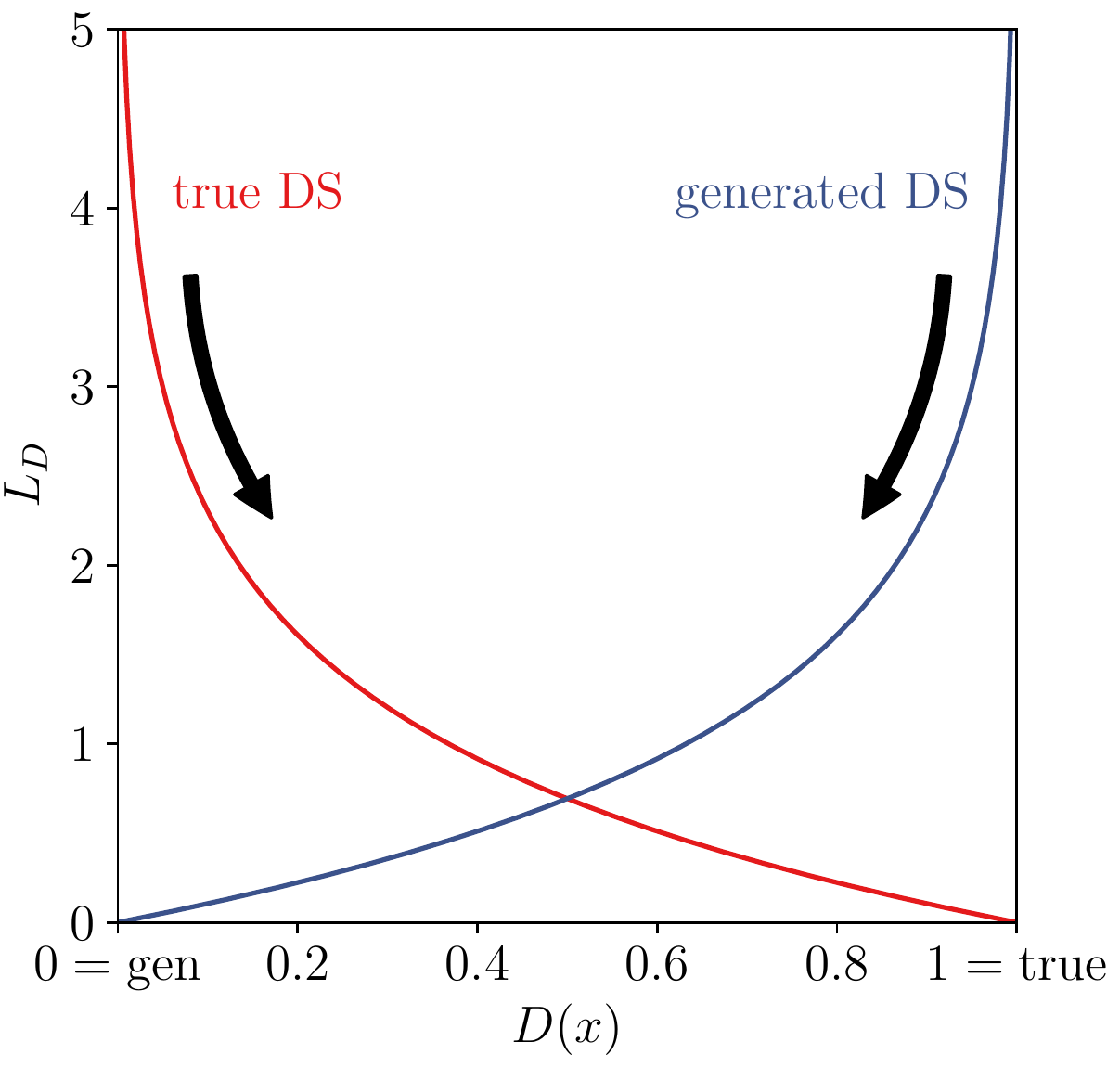}
\includegraphics[width=0.45\textwidth]{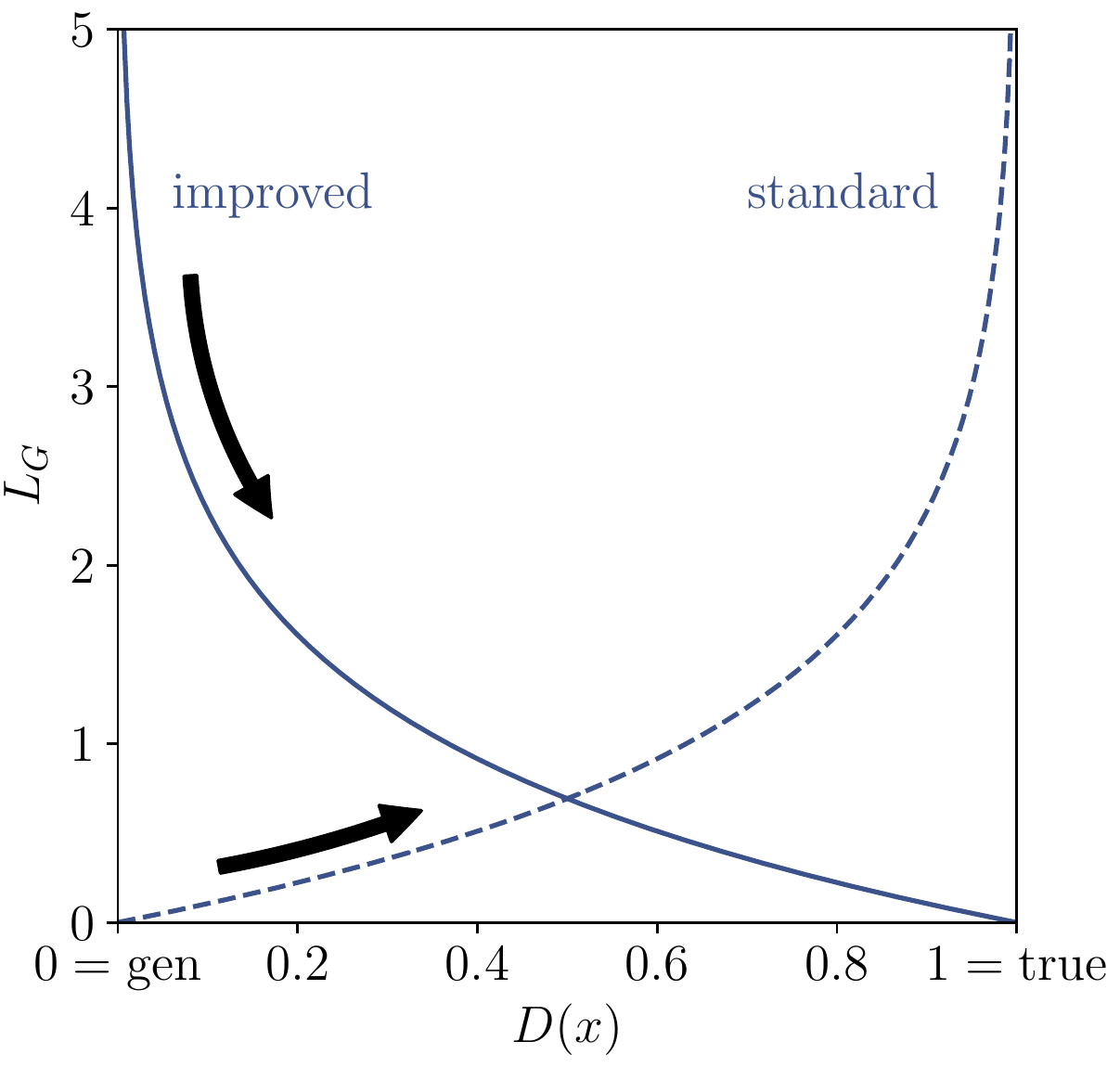}
\caption{Discriminator and generator losses as a function of the value
  assigned by the discriminator. The red line indicates batches from
  the true distribution, the blue lines batches from a generated
  distribution. The arrows indicate the direction of the training.}
\label{fig:loss}
\end{figure}

The discriminator output $D(x)\in (0,1)$ is
trained to give $D=1$ for each point in a true batch and $D=0$ for the
each point in the generated and hence not true batch. We can enhance
the sensitivity for $D\to 0$ by evaluating the variable $-\log
D(x) \in (\infty,0)$ instead of  $D(x)$ in the expectation value
\begin{align}
\Langle -\log D(x) \Rangle_x = -\frac{1}{N_x} \; \sum_{x \in \text{batch}} \log D(x) \; ,
\end{align}
where $N_x$ is the batch size. For a correctly labelled true sample
this expectation value gives zero. The loss function is defined such
that it becomes minimal when the discriminator correctly predicts the
true and generated batches
\begin{align}
L_D 
=   \Langle -\log D(x) \Rangle_{x \sim P_T} 
  + \Langle - \log (1-D(x)) \Rangle_{x \sim P_G} \; .
\label{eq:GAN1}
\end{align}
The symbol $x \sim P$ indicates phase space configurations
sampled from $P$.  In the GAN application this discriminator network
gets successively re-trained for a fixed truth $P_T(x)$ but evolving
$P_G(x)$, as illustrated in the left panel of Fig.~\ref{fig:loss}. We
can compute the discriminator loss in the limit where the generator
has produced a perfect image of the true distribution. In this case
the discriminator network will give $D=0.5$ for each point $x$ and the
result becomes $L_D = -2 \log 0.5 \approx1.4$.

The generator network starts from random noise and transforms it into
a distribution $P_G(x)$. For this it relies on the function $D(x)$,
which encodes the truth information.  Following Eq.\eqref{eq:GAN1}
this means we can maximize its second term in the training of the
generator network. It turns out that it is numerically more efficient
to instead minimize the generator loss
\begin{align}
L_G= \Langle - \log D(x) \Rangle_{x \sim P_G} \; .
\label{eq:GAN2}
\end{align}
In the right panel of Fig.~\ref{fig:loss} we see how this
assignment leads to larger gradients away from the true
configurations.

The key to the GAN training is the alternating training of the
generator and discriminator networks with their respective loss
functions given in Eq.\eqref{eq:GAN1} and Eq.\eqref{eq:GAN2}. 
Here, the balance between generator and
discriminator is crucial. On the one hand, the generator can only be
as good as the discriminator which defines the level of similarity
between true and generated data.  On the other hand, a perfect
discriminator leads to a vanishing loss function, which reduces the
gradient and slows down the training.  This interplay of the two
networks often leads to stability issues in the
training~\cite{gan_convergence}. A common way to stabilize networks
are noise-induced regularization methods, or equivalently including a
penalty on the gradient for the discriminator variable
$D(x)$~\cite{gan_stabilize_training}. Specifically, we apply the
gradient to the monotonous logit function
\begin{align}
\phi(x) = \log \frac{D(x)}{1-D(x)}
\qquad \Rightarrow \qquad 
\frac{\partial \phi}{\partial x}
= \frac{1}{D(x)} \frac{1}{1-D(x)} 
  \frac{\partial D}{\partial x}
\end{align}
enhancing its sensitivity in the regions $D \to 0$ or $D \to 1$. The
penalty applies to regions where the discriminator loss leads to a
wrong prediction, $D \approx 0$ for a true batch or $D \approx 1$
away from the truth. This means we add a term to the discriminator
loss and obtain the regularized Jensen-Shannon GAN objective\cite{gan_stabilize_training}:
\begin{align}
L_D \to 
L_D 
+ \lambda_D
\Langle \left(1- D(x)\right)^2 \vert \nabla \phi \vert^2 \Rangle_{x \sim P_T} 
+ \lambda_D
\Langle D(x)^2\, \vert \nabla \phi \vert^2 \Rangle_{x \sim P_G} \; ,
\label{eq:dloss}
\end{align}
with a properly chosen variable $\lambda_D$. The pre-factors $(1-D)^2$
and $D^2$ indeed ensure that for a properly trained discriminator this
additional contribution vanishes. Another method to avoid
instabilities in the training of the GAN is to use the Wasserstein
distance~\cite{wgan_original,wgan_gp} but our tests have shown that
Eq.\eqref{eq:dloss} works better in our case.

As a side remark, another common type of neural network used for
generative problems are variational autoencoders (VAE). They perform a
dimensional reduction of the input data --- often an image --- to
create a latent representation.  The autoencoder is trained to
minimize the difference between input and inferred image, where a
variational autoencoder requires the components of the latent
representation to follow a Gaussian.  If we then insert Gaussian
random numbers for the latent representation, the decoder generates
new images with the same characteristics as the training data. While
VAEs can be used to generate new data samples, a key component is the
latent modelling and the marginalization of unnecessary variables,
which is not a problem in generating LHC events.

\subsection{Loss functions for intermediate particles}

A particular challenge for our phase space GAN will be the
reconstruction of the $W$ and top masses from the final-state
momenta. For instance, for the top mass the discriminator and
generator have to probe a 9-dimensional part of the phase space, where
each direction covers several 100~GeV to reproduce a top mass peak
with a width of $\Gamma_t=1.5$~GeV. Following the discussion of the
Monte Carlo methods in Sec.~\ref{sec:standard} the question is how we
can build an analogue to the phase space mappings for Monte
Carlos. Assuming that we know which external momenta can form a
resonance we explicitly construct the corresponding invariant masses
and give them to the neural network to streamline the comparison
between true and generated data. We emphasize that this is
significantly less information than we use in Eq.\eqref{eq:ps_mapping},
because the network still has to learn the intermediate particle mass,
width, and shape of the resonance curve.

A suitable tool to focus on a low-dimensional part of the full phase
space is the maximum mean discrepancy (MMD)~\cite{mmd}. The MMD is a
kernel-based method to compare two samples drawn from different
distributions. Using one batch of true data points and one batch of
generated data points, it computes a distance between the distributions
as
\begin{align}
\text{MMD}^2(P_T,P_G)
&= 
\Langle  k(x, x') \Rangle_{x, x' \sim P_T}
+ \Langle  k(y, y') \Rangle_{y, y' \sim P_G}
-2 \Langle  k(x, y) \Rangle_{x \sim P_T, y  \sim P_G} \; ,
\end{align}
where $k(x,y)$ can be any positive definite kernel
function. Obviously, two identical distributions lead to
$\text{MMD}(P,P)=0$ in the limit of high statistics. Inversely, if
$\text{MMD}(P_T,P_G)=0$ for randomly sampled batches the two
distributions have to be identical $P_T(x) = P_G(x)$. The shape of the
kernels determines how local the comparison between the two
distributions is evaluated. Two examples are Gaussian or Breit-Wigner
kernels
\begin{align}
k_\text{Gauss}(x,y) =  \exp - \dfrac{(x-y)^2}{2 \sigma^2}
\qquad \text{or} \qquad 
k_\text{BW}(x,y) = \dfrac{\sigma^2}{(x-y)^2 + \sigma^2}\; ,
\label{eq:kernels}
\end{align}
where the hyperparameter $\sigma$ determines the resolution.  For an
optimal performance it should be of the same order of magnitude as the
width of the feature we are trying to learn. If the resonance and the kernel width become too narrow, we can improve convergence by
including several kernels with increasing widths to the loss
function. The shape of the kernel has nothing to do with the shape of
the distributions we are comparing. Instead, the choice between the
exponentially suppressed Gaussian and the quadratically suppressed
Breit-Wigner determines how well the MMD accounts for the tails around
the main feature.  As a machine learning version of phase space
mapping we add this MMD to the generator loss
\begin{align}
L_G \to 
L_G 
+ \lambda_G \, \text{MMD}^2 \; ,
\label{eq:gloss}
\end{align}
with another properly chosen variable $\lambda_G$.

Similar efforts in using the MMD to generate events have already been done in \cite{GMMN,MomentMatching, MMDgen} and has also been extended to a adversarial MMD version or MMD-GAN \cite{MMDGAN,DemystifyingMMDGAN, MMDGANIKL}, in which the MMD kernel is learned by another network.

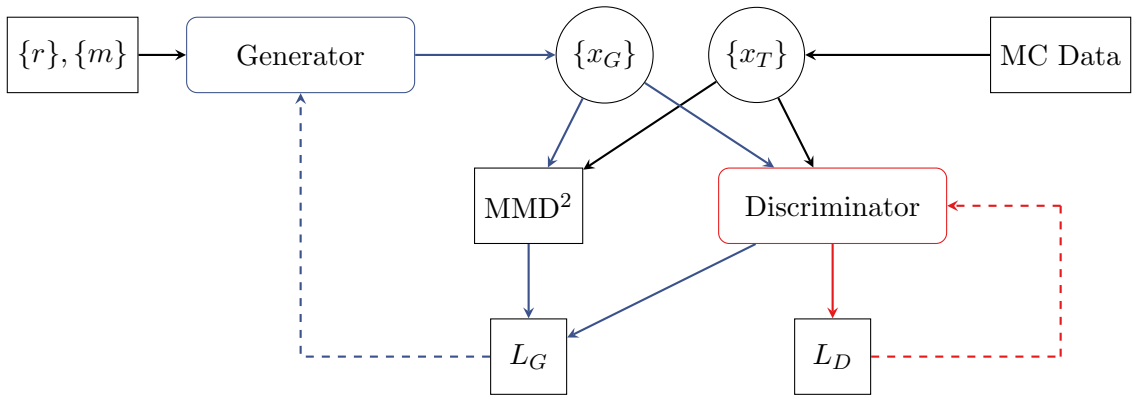
\begin{figure}[t]
\centering
\begin{tikzpicture}[node distance=2cm]

\node (generator) [generator] {Generator};
\node (random) [process,left of=generator, xshift=-1cm] {$\{ r \}, \{ m \}$};
\draw [arrow, color=black] (random) -- (generator);

\node (in1) [io, right of=generator, xshift=2cm] {$\{ x_G \}$};
\draw [arrow,color=Gcolor] (generator) -- (in1);

\node (in2) [io, right of=in1] {$\{ x_T \}$};
\node (data) [process, right of=in2, xshift=2cm] {MC Data};
\draw [arrow,color=black] (data) -- ( in2);

\node (discriminator) [discriminator, below of=in2,xshift=1cm] {Discriminator};
\node (mmd) [process, below of=in1,xshift=-1cm] {MMD$^2$};

\draw [arrow, color=Gcolor] (in1) -- (mmd);
\draw [arrow] (in2) -- (mmd);
\draw [arrow, color=Gcolor] (in1.325) -- (discriminator);

\draw [arrow, color=black] (in2) -- (discriminator);

\node (gloss) [decision, below of=mmd] {$L_G$};
\draw [arrow,color=Gcolor] (mmd) -- (gloss);
\node (dloss) [decision, below of=discriminator] {$L_D$};
\draw [arrow,color=Dcolor] (discriminator) -- (dloss);
\draw [arrow,color=Gcolor] (discriminator) -- (gloss);

\draw [arrow,dashed, color=Gcolor] (gloss) -| (generator);
\coordinate[right of=dloss, xshift = 1cm](a);
\coordinate[above of=a](b);

\draw[thick,dashed,color=Dcolor] (dloss) -- (a);
\draw[thick,dashed,color=Dcolor] (a) -- (b);
\draw[arrow,dashed, color=Dcolor] (b) -- (discriminator);

\end{tikzpicture}
\caption{Schematic diagram for our GAN.  The input $\{r\}$ and $\{
  m\}$ describe a batch of random numbers and the masses of the
  external particles, and $\{ x \}$ denotes a batch of phase space
  points sampled either from the generator or the true data.  The blue
  (red) and arrows indicate which connections are used in the training
  of the generator (discriminator).}
\label{fig:GAN}
\end{figure}

In Fig.~\ref{fig:GAN} we show the whole setup of our network. It works
on batches of simulated parton-level events, or unweighted event
configurations $\{ x\}$. The input for the generator are batches of
random numbers $\{ r\}$ and the masses $\{ m\}$ of the final state
particles. Because of the random input a properly trained GAN will
generate statistically independent events reflecting the learned
patterns of the training data. For both the generator and the
discriminator we use a 10-layer MLP with 512 units each, the remaining
network parameters are given in Tab.~\ref{tab:details}. The main
structural feature of the competing networks is that the output of the
discriminator, $D$, is computed from the combination of true and
generated events and is needed by the generator network. The generator
network combines the information from the discriminator and the MMD in
its loss function, Eq.\eqref{eq:gloss}. The learning is done when the
distribution of generated unweighted events $\{ x_G \}$ and true
Monte-Carlo events $\{ x_T \}$ are essentially identical. We again
emphasize that this construction does not involve weighted events.

\begin{table}[t]
\begin{small} \begin{center}
\begin{tabular}{l r}
\toprule
Parameter              & Value  \\
\midrule
Input dimension G & $18+6$\\
Layers & 10 \\
Units per layer & 512 \\
Trainable weights G & \quad 2382866\\
Trainable weights D & 2377217\\
\midrule
$\lambda_D$ & $10^{-3}$\\
$\lambda_G$ & 1 \\
\midrule
Batch size & 1024 \\
Epochs & 1000\\
Iterations per epoch & 1000\\
Training time & 26h \\
Size of trainings data & $10^6$\\
\bottomrule
\end{tabular}
\end{center} \end{small}
\caption{Details for our GAN setup.}
\label{tab:details}
\end{table}

\section{Machine-learning top pairs}
\label{sec:ps_gan}

A sample Feynman diagram for our benchmark process 
\begin{align}
pp \to t \bar{t} \to (b q \bar{q}') \; (\bar{b} \bar{q} q') 
\end{align}
is shown in Fig.~\ref{fig:feyn_intro}.  For our analysis we generate
1~million samples of the full $2\to6$ events as training data sample with
\madgraph\cite{madgraph}.  The intermediate tops and $W$-bosons allow
us to reduce the number of Feynman diagrams by neglecting proper
off-shell contributions and only including the approximate
Breit-Wigner propagators. Our results can be directly extended to a
proper off-shell
description~\cite{offshell_worek,offshell_gudrun,offshell_ansgar},
which only changes the details of the subtle balance in probing small
but sharp on-shell contributions and wide but flat off-shell
contributions. Similarly, we do not employ any detector simulation,
because this would just wash out the intermediate resonances and
diminish our achievement unnecessarily.

Because we do not explicitly exploit momentum conservation our final
state momenta are described by 24 degrees of freedom. Assuming full
momentum conservation would for instance make it harder to include
approximate detector effects.  These 24 degrees of freedom can be
reduced to 18 when we require the final-state particles to be
on-shell.  While it might be possible for a network to learn the
on-shell conditions for external particles, we have found that
learning constants like external masses is problematic for the GAN
setup.  Instead, we use on-shell relations for all final-state momenta
in the generator network.

Combining the GAN with the MMD loss function of Eq.\eqref{eq:gloss}
requires us to organize the generator input in terms of momenta of
final-state particles. With the help of a second input to the
generator, namely a 6-dimensional vector of constant final-state
masses, we enhance the 18-dimensional input to six 4-vectors. This way
we describe all final-state particles, denoted as $\{ x_G \}$ in
Fig.~\ref{fig:GAN}, through an array
\begin{align}
x = \{p_1,p_2,p_3,p_4,p_5,p_6\}\:,
\end{align}
where we fix the order of the particles within the events. This format
corresponds to the generated unweighted truth events $\{ x_T \}$ from
standard LHC event simulators. In particular, we choose the
momenta such that
\begin{align}
p_{W^-} &= p_1+p_2\:,&
p_{W^+} &= p_4+p_5\:, &
p_{\bar{t}} &= p_1+p_2+p_3\:,&
p_t &= p_4+p_5+p_6\:.
\end{align}
For the on-shell states we extract the resonances from the full phase
space and use those to calculate the MMD between the true and the
generated mass distributions.  This additional loss is crucial to
enhance the sensitivity in certain phase space regions allowing the
GAN to learn even sharp feature structures.

\subsubsection*{Flat distributions}

\begin{figure}[t]
\centering
\includegraphics[width=0.49\textwidth]{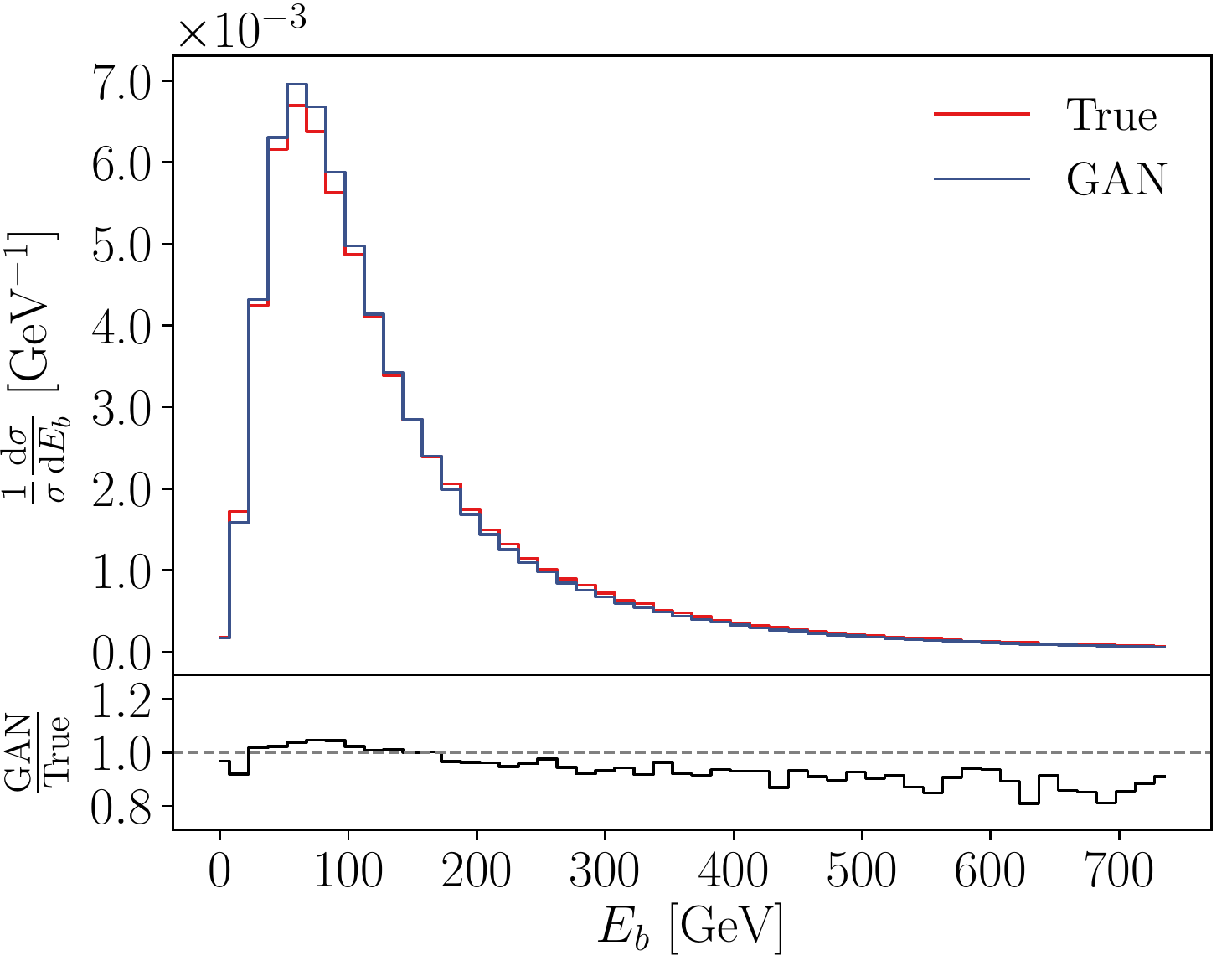}
\includegraphics[width=0.49\textwidth]{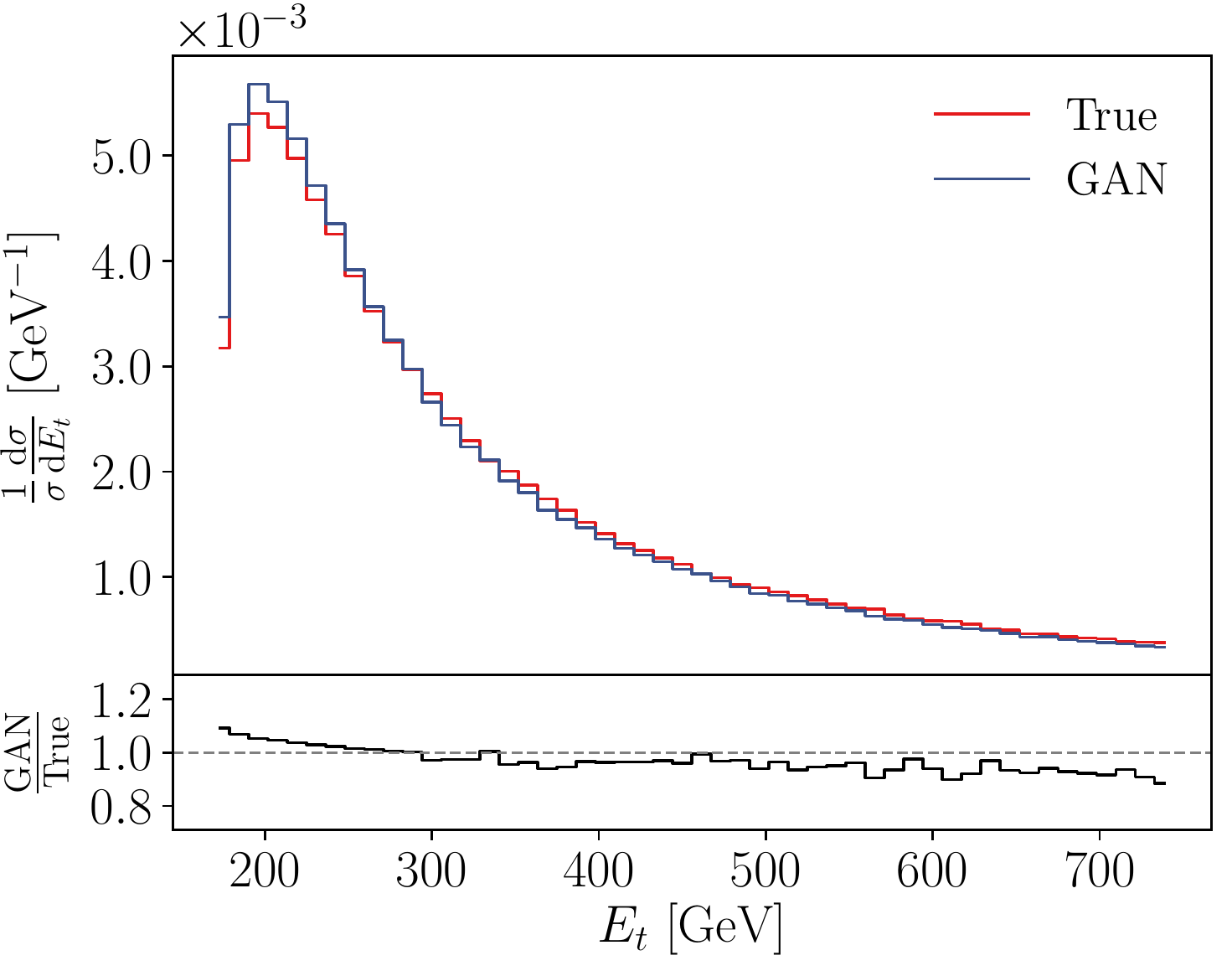}\\
\includegraphics[width=0.49\textwidth]{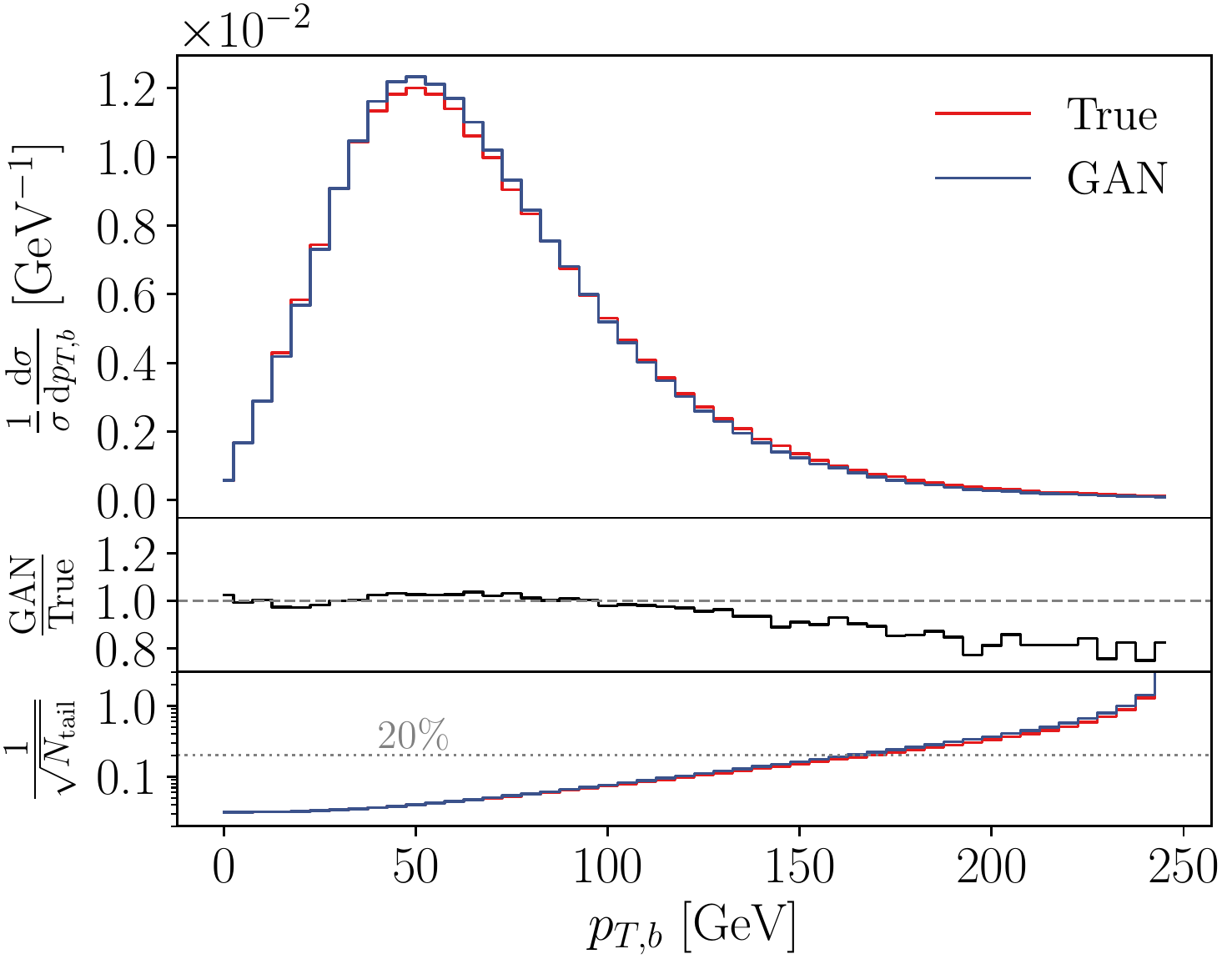}
\includegraphics[width=0.49\textwidth]{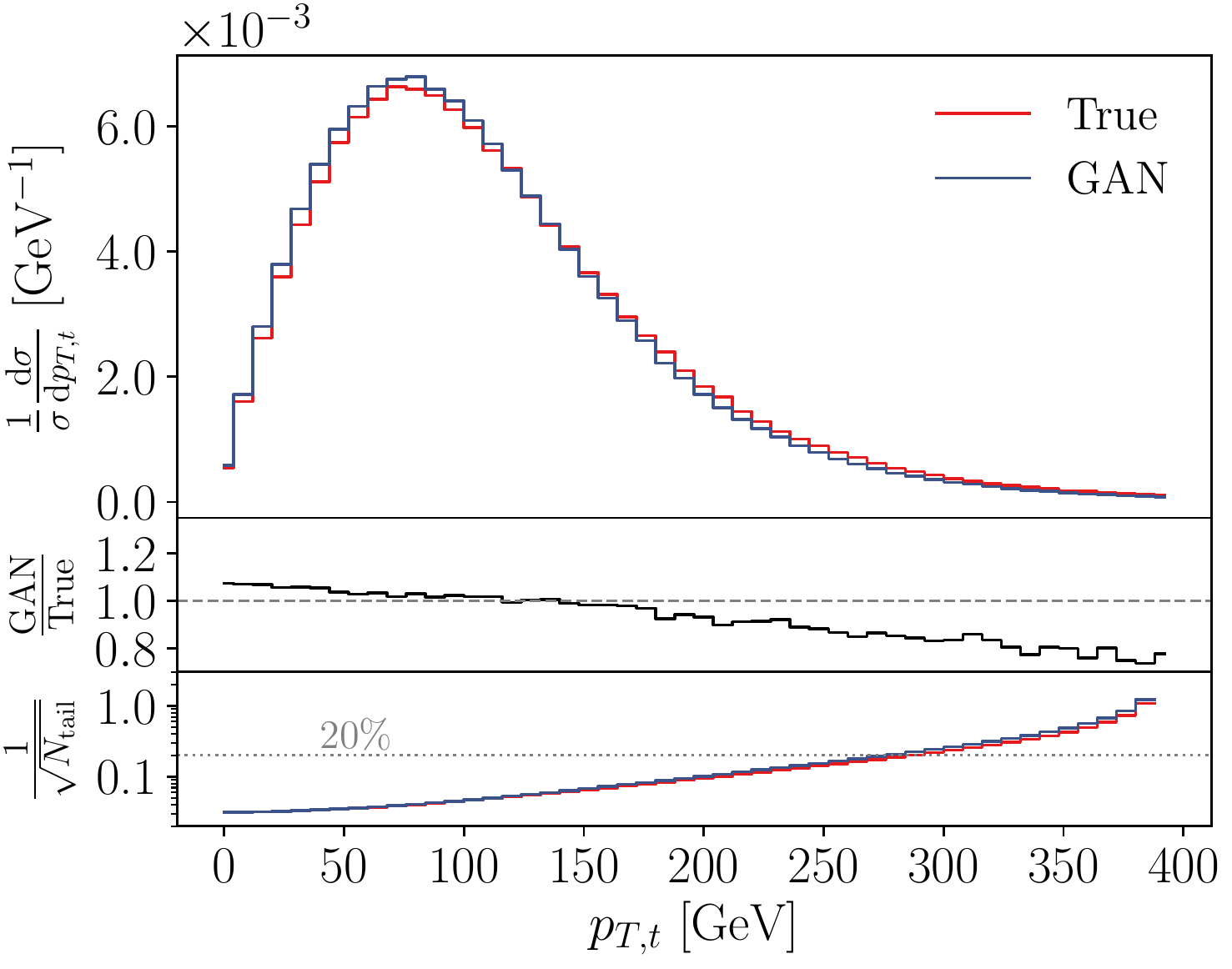}
\caption{Energy (top) and transverse momentum (bottom) distributions
  of the final-state $b$-quark (left) and the decaying top quark
  (right) for MC truth (blue) and the GAN (red). The lower panels give
  the bin-wise ratio of MC truth to GAN distribution. For the
  $p_T$~distributions we show the relative statistic uncertainty on
  the cumulative number of events in the tail of the distribution for
  our training batch size.}
\label{fig:momentum_distributions}
\end{figure}

To begin with, relatively flat distributions like energies, transverse momenta,
or angular correlations should not be hard to GAN~\cite{dutch,DijetGAN,gan_datasets}. As
examples, we show transverse momentum and energy distributions of the final-state
$b$-quarks and the intermediate top quarks in
Fig.~\ref{fig:momentum_distributions}. The GAN reproduces the true
distributions nicely even for the top quark, where the generator needs
to correlate the four-vectors of three final-state particles.  

To better judge the quality of the generator output we show the ratio
of the true and generated distributions in the lower panels of each
plot, for instance $E_b^{(G)}/E_b^{(T)}$ where $E_b^{(G,T)}$ is
computed from the generated and true events, respectively.  The
bin-wise difference of the two distributions increases to around 20\%
only in the high-$p_T$ range where the GAN suffers from low statistics
in the training sample.  To understand this effect we also quantify
the impact of the training statistics per batch for the two
$p_T$-distributions. In the set of third panels we show the relative
statistical uncertainty on the number of events $N_\text{tail}(p_T)$
in the tail above the quoted $p_T$ value.
The relative statistical uncertainty on this number of events is
generally given by $1/\sqrt{N_\text{tail}}$.  For the
$p_{T,b}$-distribution the GAN starts deviating at the 10\% level
around 150~GeV. Above this value we expect around 25 events per batch,
leading to a relative statistical uncertainty of 20\%. The top
kinematics is harder to reconstruct, leading to a stronger impact from
low statistics. Indeed, we find that the generated distribution
deviates by 10\% around $p_{T,t}\gtrsim 250$~GeV where the relative
statistic uncertainty reaches 15\%.


We emphasize that this limitation through training statistics is
expected and can be easily corrected for instance by slicing the
parameter in $p_T$ and train the different phase space regions
separately. Alternatively, we can train the GAN on events with a
simple re-weighting, for example in $p_T$, but at the expense of
requiring a final unweighting step.

\subsection*{Phase space coverage}

\begin{figure}[t]
\begin{minipage}{\textwidth}
\centering
\includegraphics[width=0.49\textwidth]{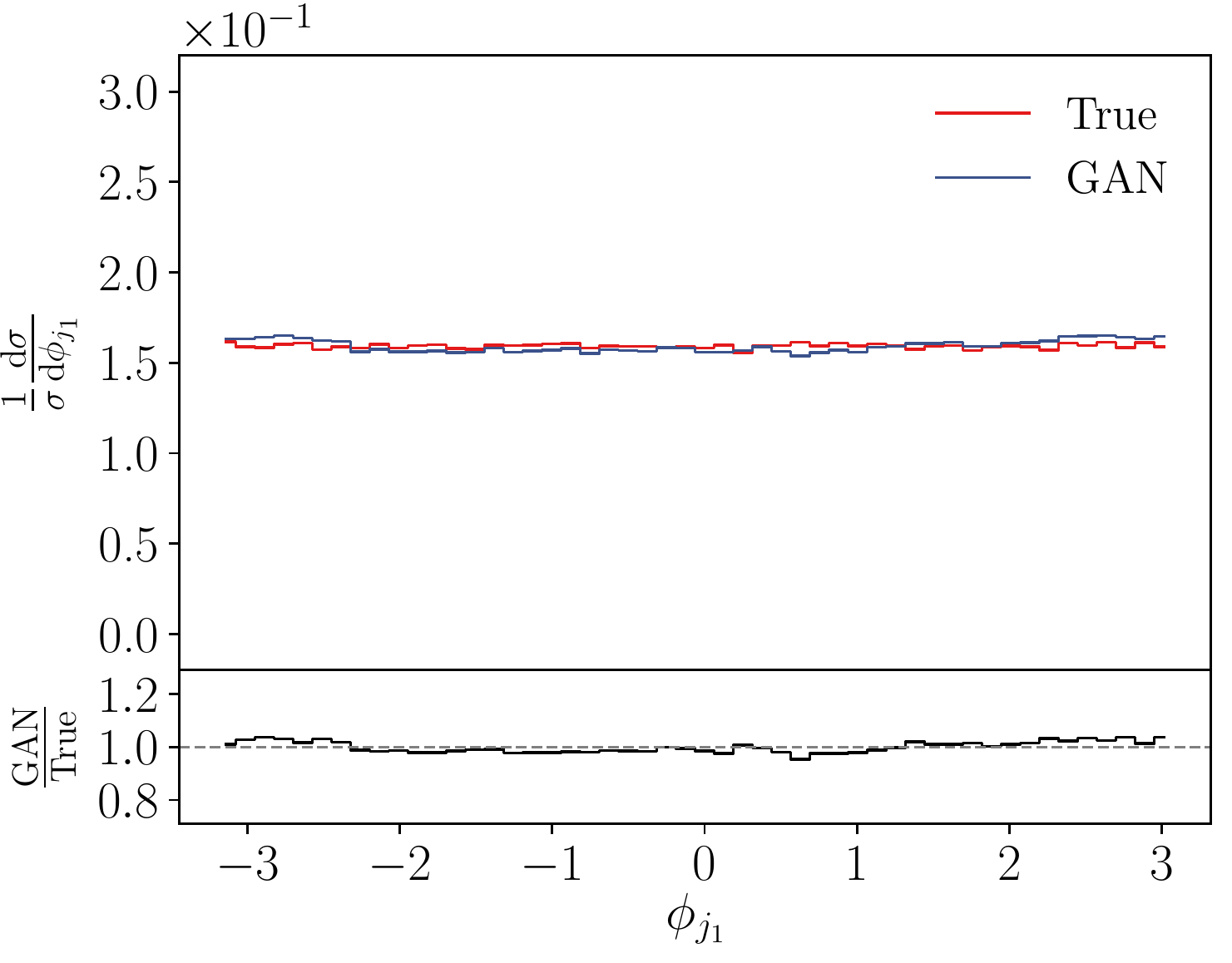}
\includegraphics[width=0.49\textwidth]{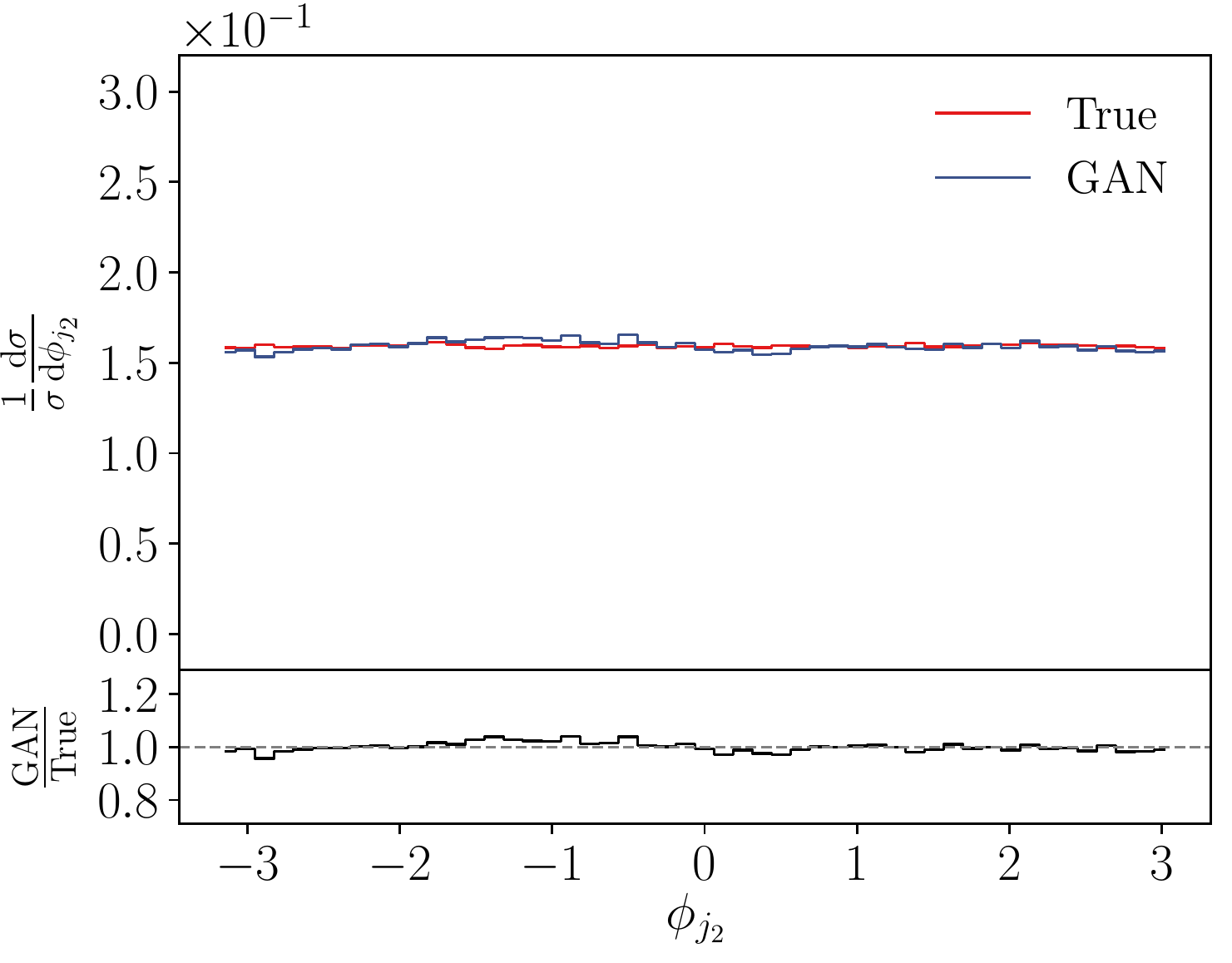} 
\caption{$\phi$ distributions of $j_1$ and $j_2$. The lower
	panels give the bin-wise ratio of MC truth to GAN
	distribution.}
\label{fig:phi-distributions}
\end{minipage}
\vspace{2em}\\
\begin{minipage}{\textwidth}
\centering
\includegraphics[width=0.49\textwidth]{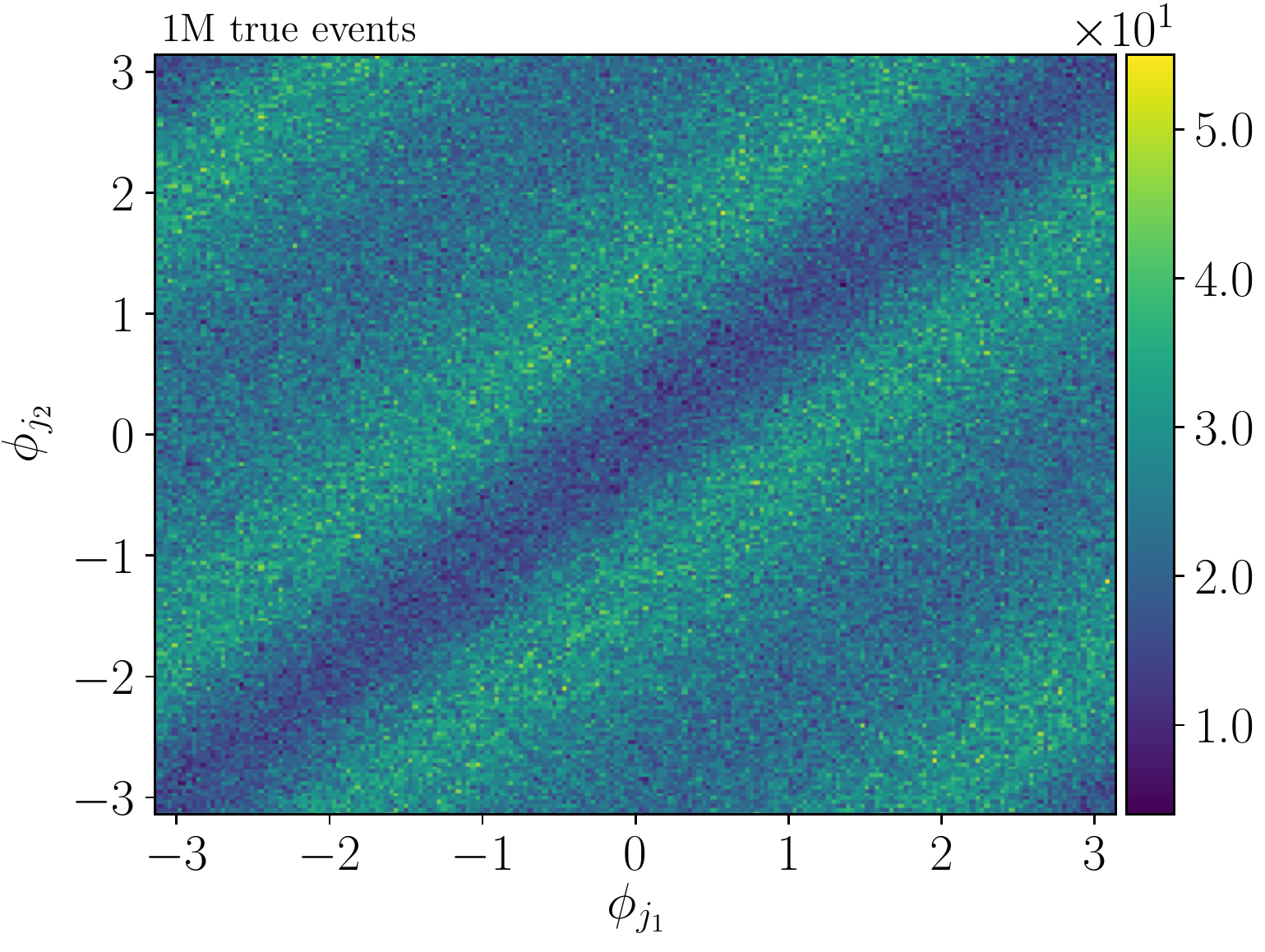}
\includegraphics[width=0.49\textwidth]{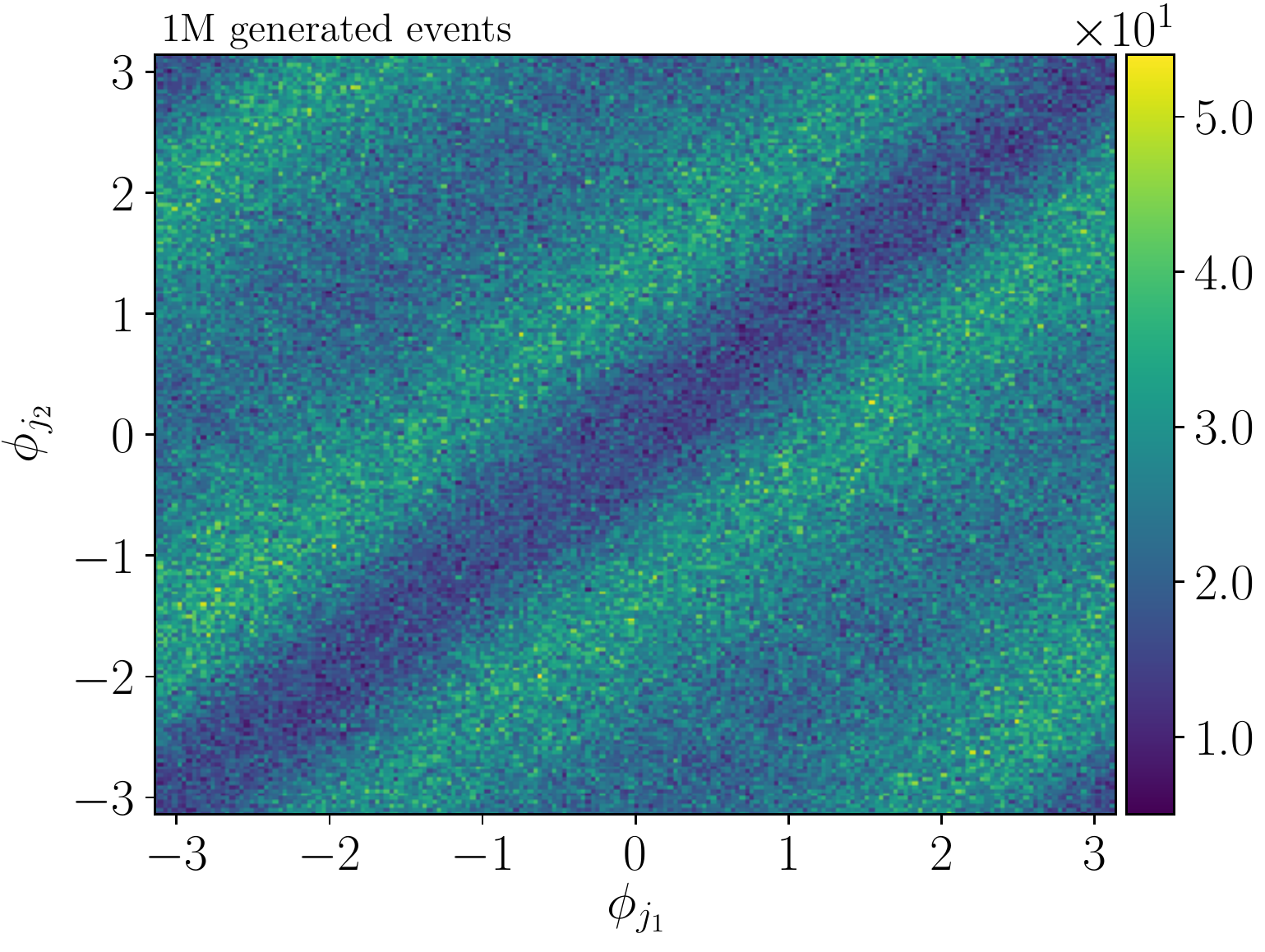}\\
\includegraphics[width=0.49\textwidth]{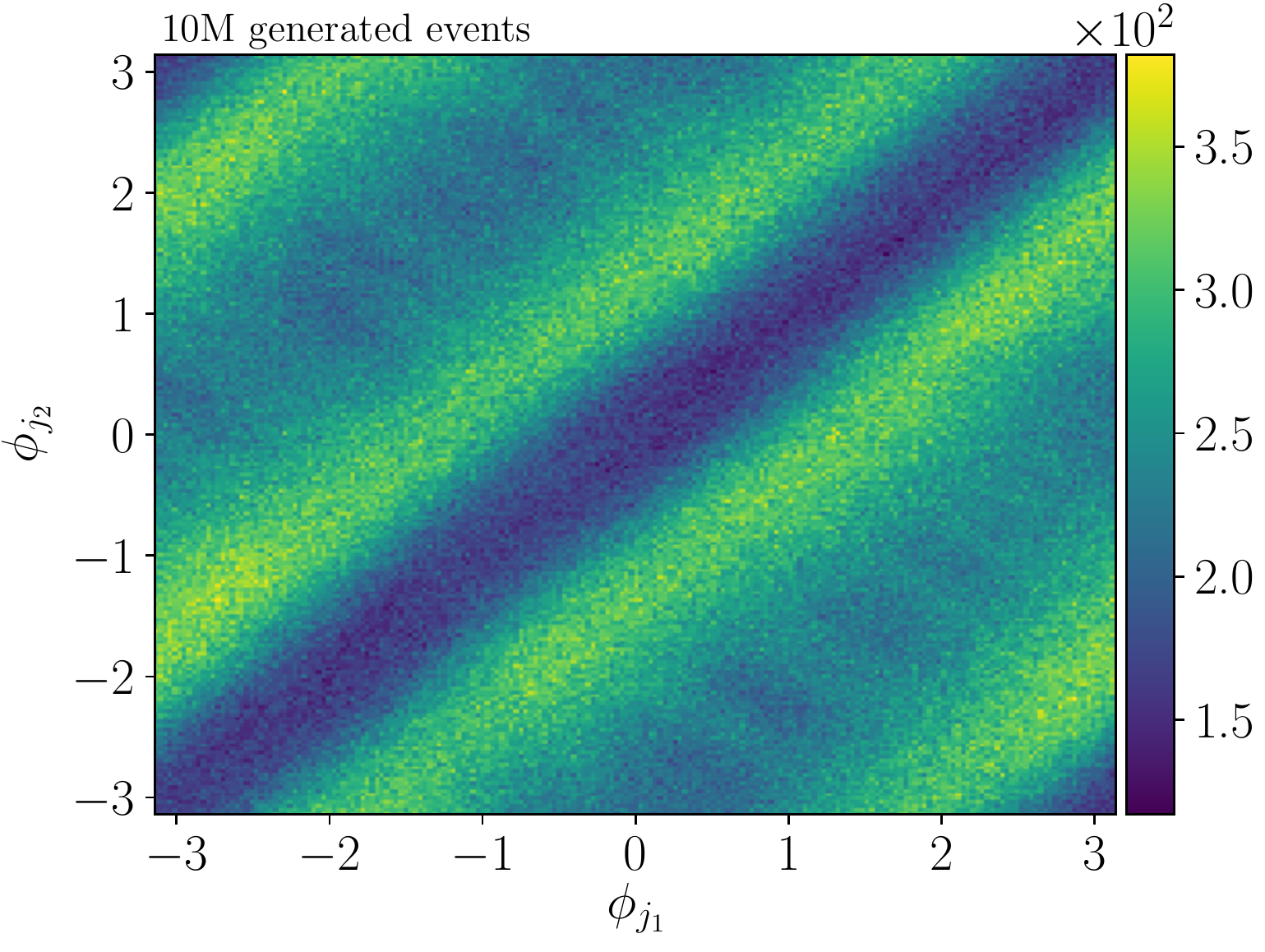}
\includegraphics[width=0.49\textwidth]{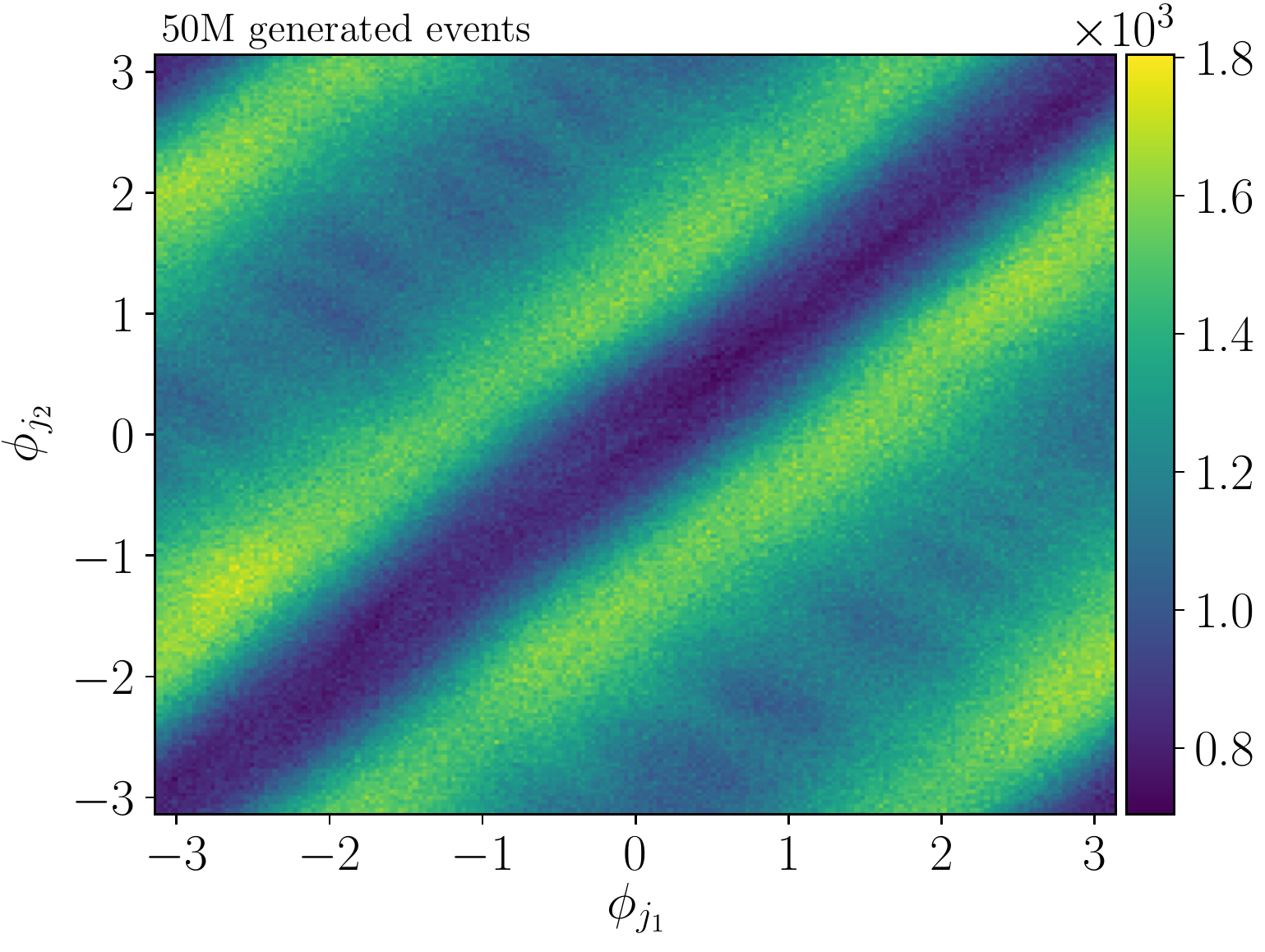}\\
\caption{Correlation between $\phi_{j_1}$ and
	$\phi_{j_2}$ for 1~million true events (upper left) and for
	1~million, 10~million, and 50~million GAN events.}
\label{fig:phi-correlations}
\end{minipage}
\end{figure}

To illustrate that the GAN populates the full phase space we can for
instance look at the azimuthal coordinates of two final-state jets in Fig.~\ref{fig:phi-distributions}. Indeed, the
generated events follow the expected flat distribution and correctly
match the true events.

Furthermore, we can use these otherwise not very interesting angular
correlations to illustrate how the GAN interpolates and generates
events beyond the statistics of the training data. In Fig.~\ref{fig:phi-correlations} we show the 2-dimensional
correlation between the azimuthal jet angles $\phi_{j_1}$ and
$\phi_{j_2}$.  The upper-left panel includes 1~million training
events, while the following three panels show an increasing number of
GANed events, starting from 1~million events up to 50~million
events. As expected, the GAN generates statistically independent
events beyond the sample size of the training data and of course
covers the entire phase space.

\subsubsection*{Resonance poles}

\begin{figure}[t]
	\centering
	\includegraphics[width=0.49\textwidth]{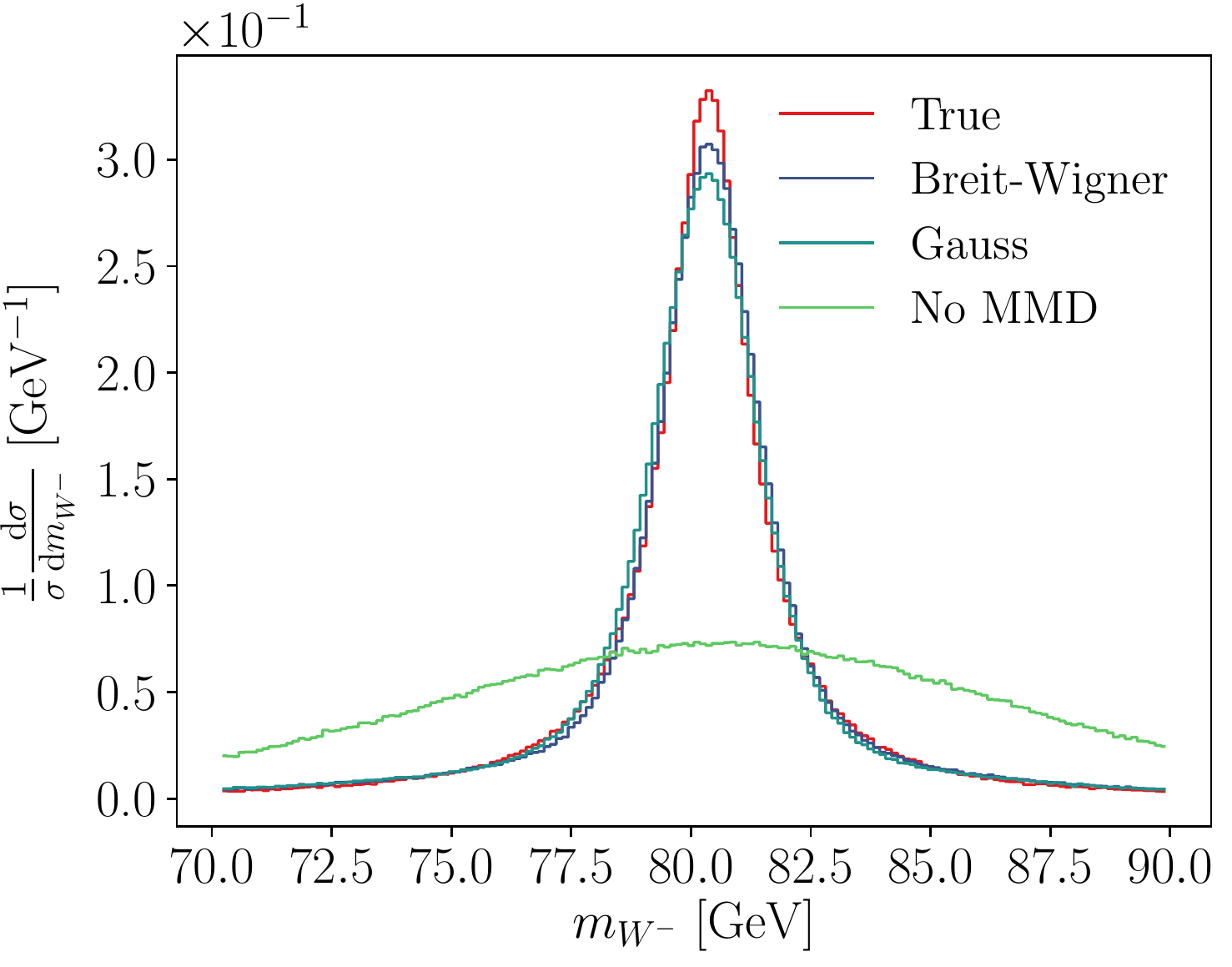}
	\includegraphics[width=0.49\textwidth]{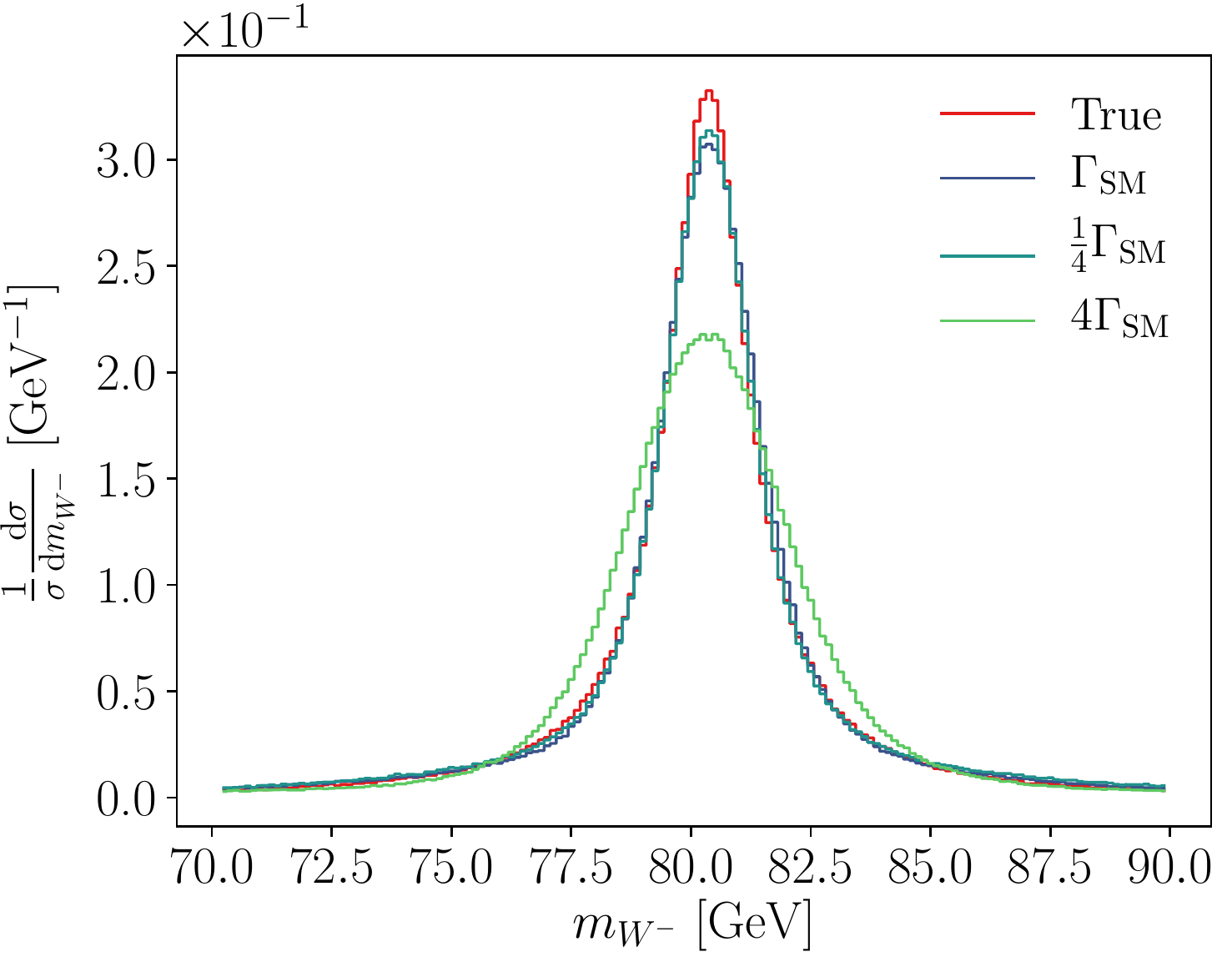}\\
	\includegraphics[width=0.49\textwidth]{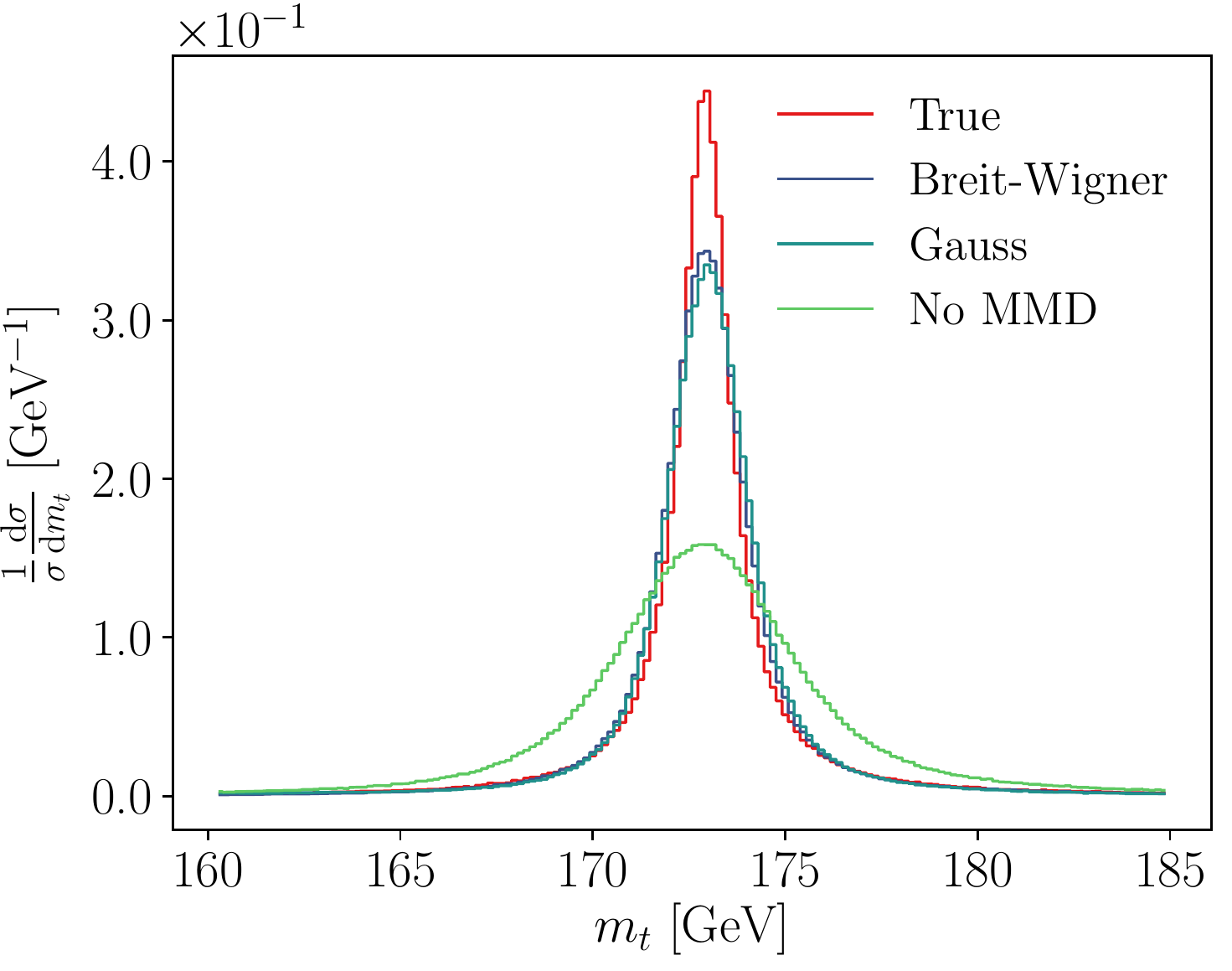}
	\includegraphics[width=0.49\textwidth]{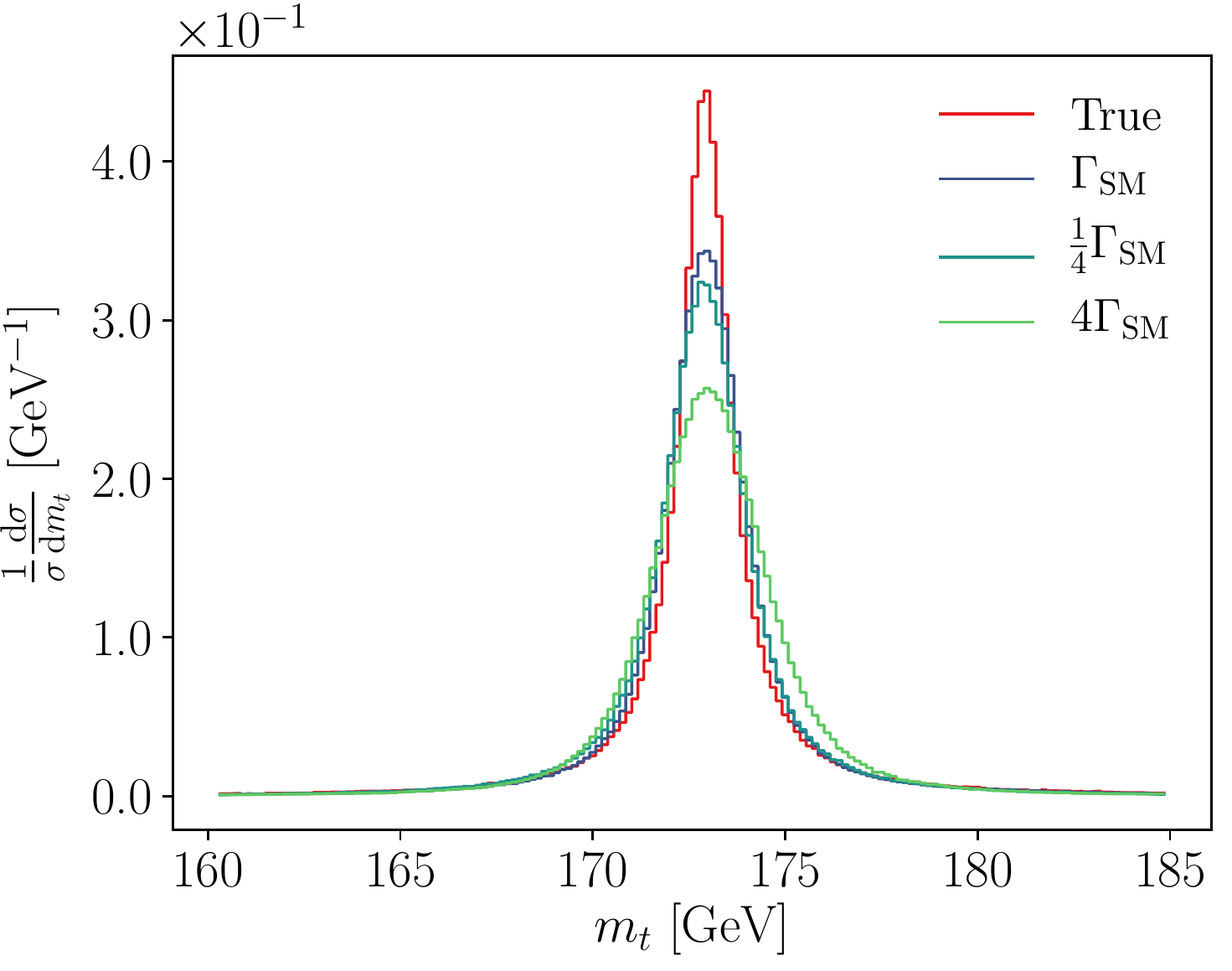}
	\caption{Comparison of different kernel functions (left) and varying widths
		(right) and their impact on the invariant mass of W boson (top) and top quark (bottom).}
	\label{fig:invariant_mass}
\end{figure}

From Ref.~\cite{maxim} we know that exactly mapping on-shell poles and
tails of distributions is a challenge even for simple decay
processes. Similar problems can be expected to arise for phase space
boundaries, when they are not directly encoded as boundaries of the
random number input to the generator.  Specifically for our $t\bar{t}$
process, Ref.~\cite{dutch} finds that their GAN setup does not
reproduce the phase space structure.  The crucial task of this paper
is to show how well our network reproduces the resonance structures of
the intermediate narrow resonances. In Fig.~\ref{fig:invariant_mass}
we show the effect of the additional MMD loss on learning the
invariant mass distributions of the intermediate $W$ and top
states. Without the MMD, the GAN barely learns the correct mass value,
in complete agreement with the findings of Ref.~\cite{gan_datasets}.
Adding the MMD loss with default kernel widths of the Standard Model
decay widths drastically improves the results, and the mass
distribution almost perfectly matches the true distribution in the
$W$-case. For the top mass and width the results are slightly worse,
because its invariant mass needs to be reconstructed from three
external particles and thus requires the generator to correlate more
variables. This gets particularly tricky in our scenario, where the
$W$-peak reconstruction directly affects the top peak. We can further
improve the results by choosing a bigger batch size as this naturally
enhances the power of the MMD loss. However, bigger batch sizes leads
to longer training times and bigger memory consumption. In order to
keep the training time on responsible level, we limited our batch size
to 1024 events per batch. As already pointed out, the results are not
perfect in this scenario, especially for the top invariant mass,
however, we can clearly see the advantages of adding the MMD loss.

To check the sensitivity of the kernel width on the results, we vary
it by factors of $\{1/4, 4\}$. As can be seen in the lower
panels of both distributions, increasing the resolution of the kernel
or decreasing the kernel width hardly affects the network
performance. On the other hand, increasing the width decreases the
resolution and leads to too broad mass peaks.  Similarly, if we switch
from the default Breit-Wigner kernel to a Gaussian kernel with the
same width we find identical results.  This means that the only thing
we need to ensure is that the kernel can resolve the widths of the 
analyzed features.

We emphasize again that we do not give the GAN the masses or even
widths of the intermediate particles. This is different from
Ref.~\cite{gan_datasets}, which tackles a similar problem for the $Z
\to \ell \ell$ resonance structure and uses an explicit mass-related
term in the loss function. We only specify the two final-state momenta
for which the invariant mass can lead to a sharp phase space structure
like a mass peak, define a kernel like those given in
Eq.\eqref{eq:kernels} with sufficient resolution and let the GAN do
the rest. This approach is even more hands-off than typical phase
space mappings employed by standard Monte Carlos.

\subsubsection*{Correlations}

\begin{figure}[t]
	\centering
	\includegraphics[width=0.49\textwidth]{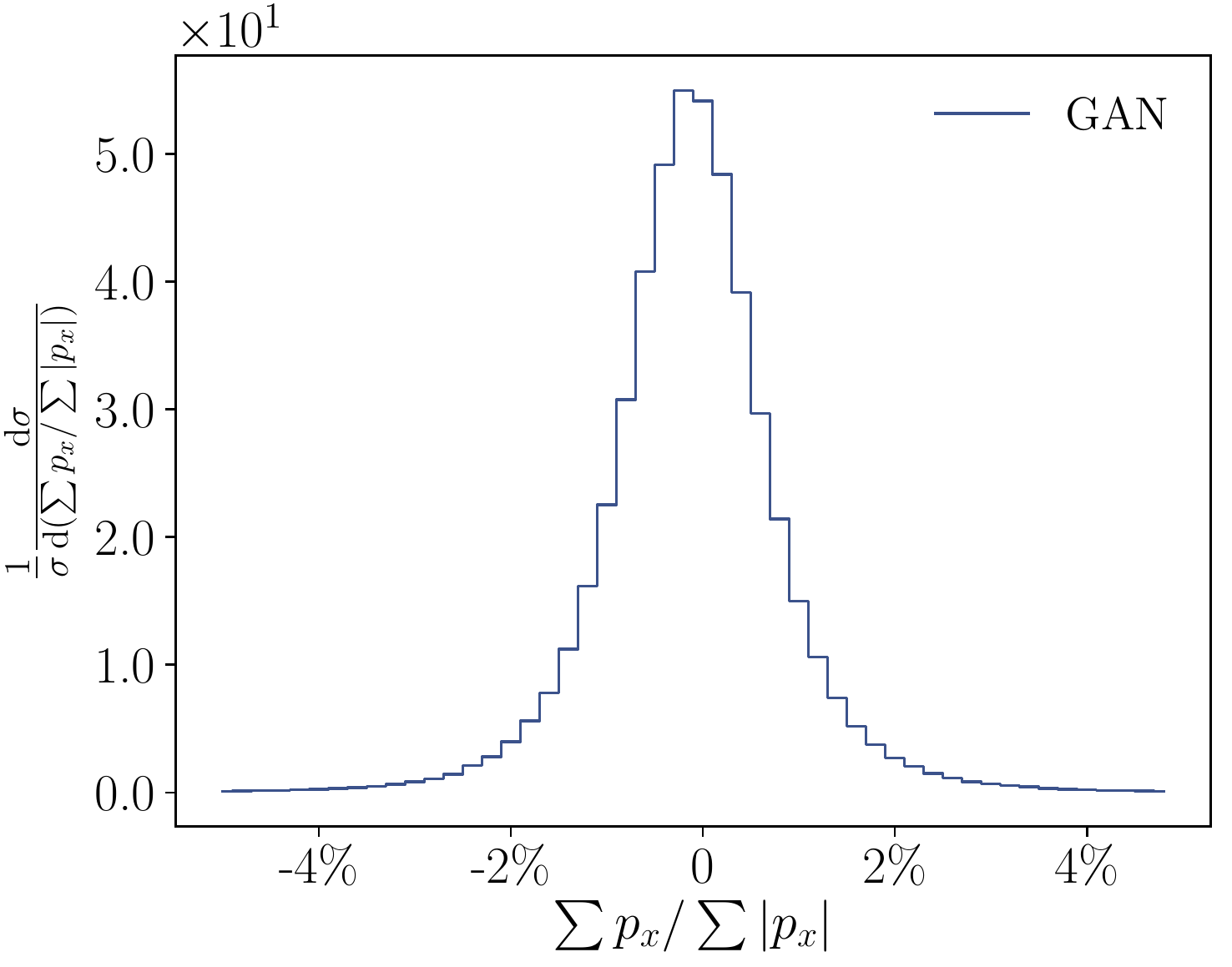}
	\includegraphics[width=0.49\textwidth]{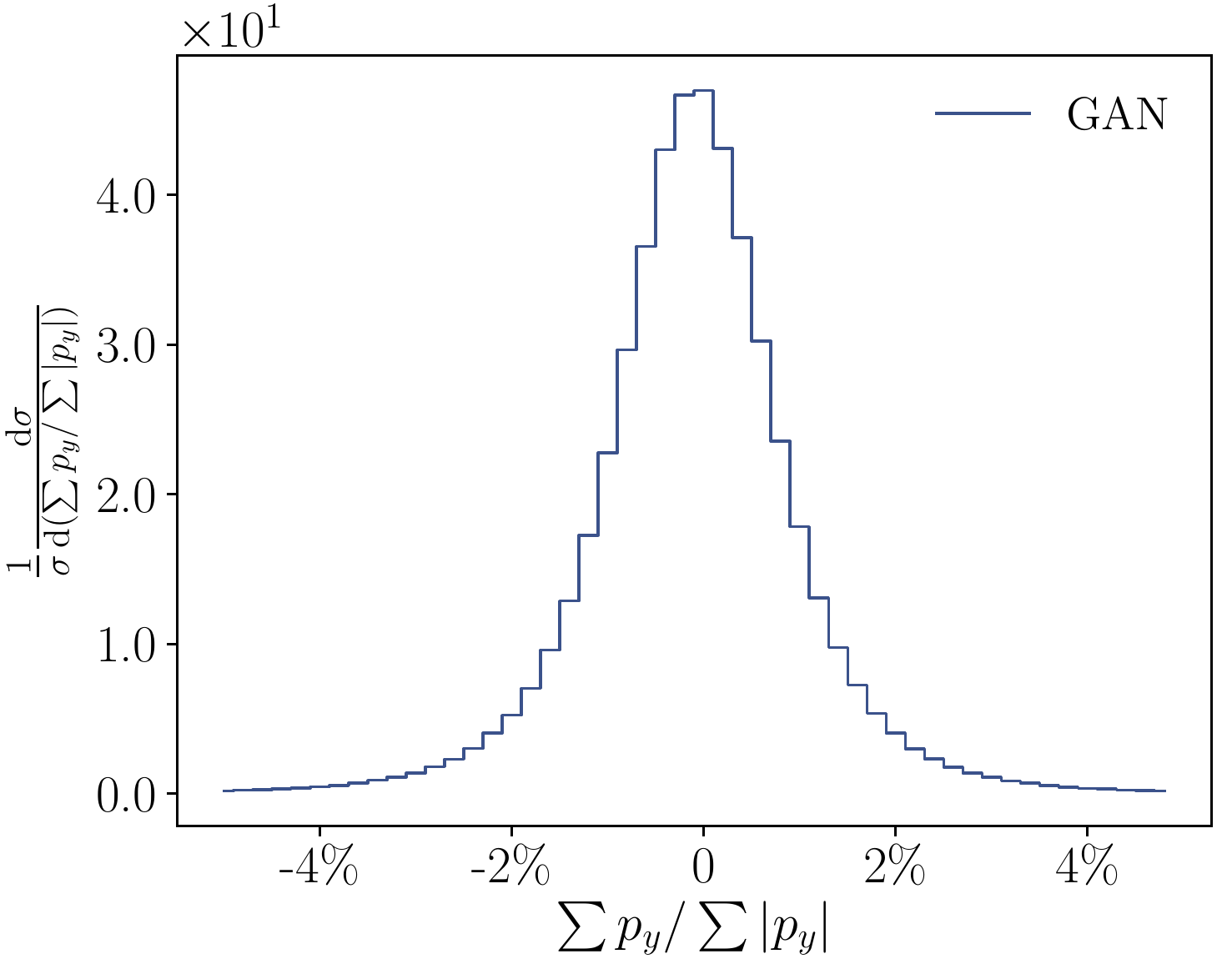}
	\caption{Sum of all $p_x$ ($p_y$) momenta divided by the sum of the
		absolute values in the left (right) panel, testing how well the GAN
		learns momentum conservation.}
	\label{fig:mom_conserve}
\end{figure}

\begin{figure}[t]
	\centering
	\centering
	\includegraphics[width=0.98\textwidth]{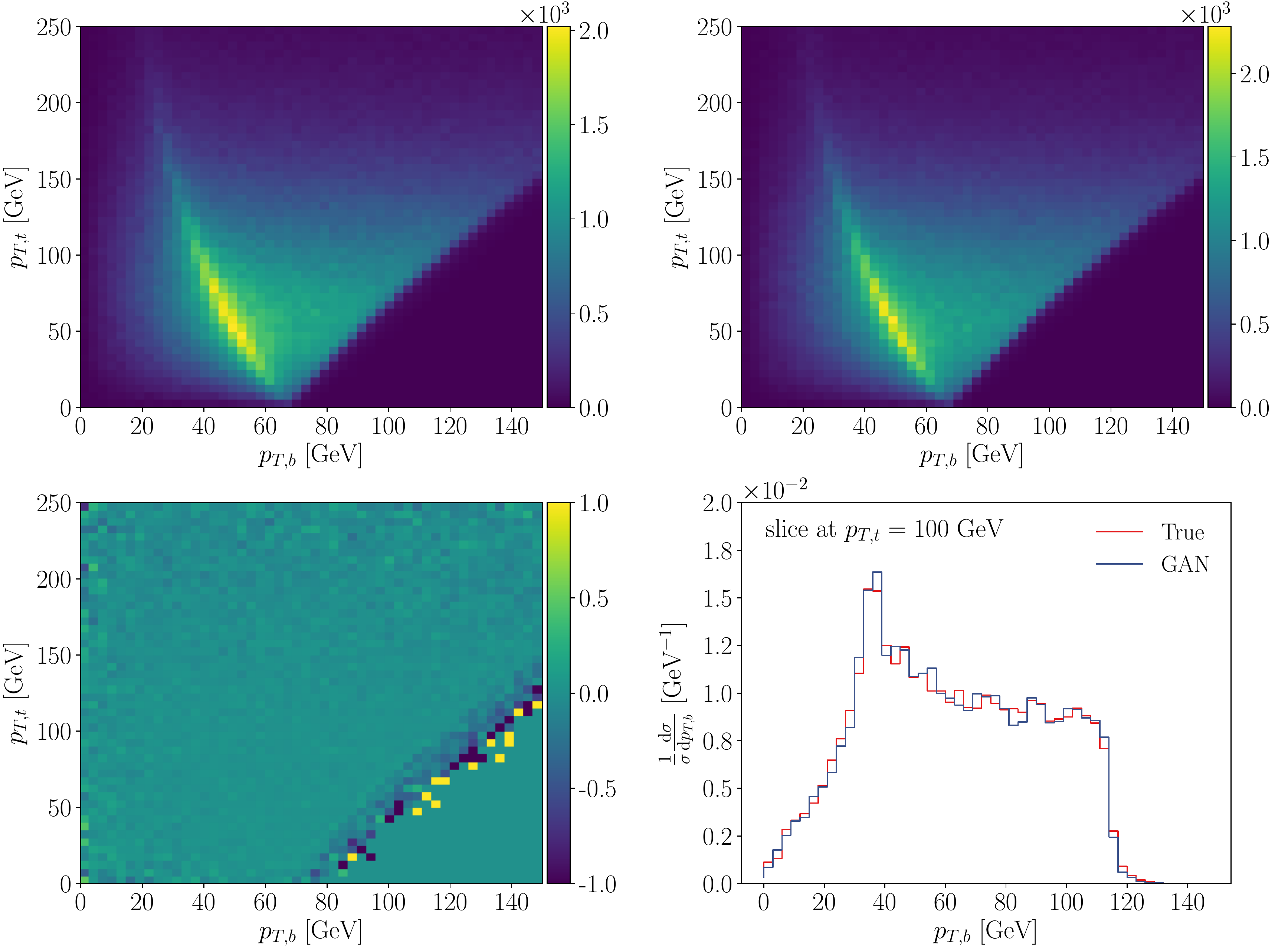}
	\caption{Correlation between $p_{T,t}$ and $p_{T,b}$ for the true data
		(upper left), GAN data (upper right) and the asymmetry between both (lower
		left). In addition, we show $p_{T,b}$ sliced at $p_{T,t}=100 \pm 1$~GeV
		(lower right).}
	\label{fig:correlations}
\end{figure}

Now that we can individually GAN all relevant phase space structures
in top pair production, it remains to be shown that the network also
covers all correlations. A simple test is 4-momentum conservation,
which is not guaranteed by the network setup.  In
Fig.~\ref{fig:mom_conserve}, we show the sums of the transverse
components of the final-state particle momenta divided by the sum of
their absolute values. As we can see, momentum conservation at the GAN
level is satisfied at the order of 2\%.

Finally, in Fig.~\ref{fig:correlations} we show 2-dimensional
correlations between the transverse momenta of the outgoing $b$-quark
and the intermediate top for the true (left) and GAN events
(right). The phase space structure encoded in these two observables is
clearly visible, and the GAN reproduces the peak in the low-$p_T$
range, the plateau in the intermediate range, and the sharp boundary
from momentum conservation in the high-$p_T$ range.  To allow for a
quantitative comparison of true and generated events we show the
bin-wise asymmetry in the lower left panel. Except for the phase space
boundary the agreement is essentially perfect. The asymmetry we
observe along the edge is a result from very small statistics. For an
arbitrarily chosen $p_T$ value of 100~GeV the deviations occur for
$p_{T,b} \in [130, 140]$~GeV.  We compare this region of statistical
fluctuations in the asymmetry plot with a 1-dimensional slice of the
correlation plot (lower right) for $p_{T,t}=100 \pm 1$~GeV. The
1-dimensional distributions shows that in this range the normalized
differential cross section has dropped below the visible range.

\section{Outlook}

We have shown that it is possible to GAN the full phase space
structure of a realistic LHC process, namely top pair production all
the way down to the kinematics of the six top decay jets. Trained 
on a simulated set of unweighted events this allows us to generate 
any number of new events representing the same phase space information. With the
help of an additional MMD kernel we described on-shell resonances as
well as tails of distributions. The only additional input was the
final-state momenta related to on-shell resonances, and the rough
phase space resolution of the on-shell pattern.

Our detailed comparison showed that relatively flat distributions can
be reproduced at arbitrary precision, limited only by the statistics
of the training sample. The mass values defining intermediate
resonance poles were also easily extracted from the dynamic GAN
setup. Learning the widths of the Breit-Wigner propagator requires an
MMD kernel with sufficient resolution and is in our case only limited
by the training time.  The main limitation of the GAN approach is that
statistical uncertainties in poorly populated tails of distributions
in the training data appear as systematic uncertainties in the same
phase space regions for the generated high-statistics samples. We have
studied this effect in detail.

Because such a GAN does not require any event unweighting we expect it
to be a useful and fast\footnote{Once trained, our GAN generates 1 million events in 1.6 minutes on a laptop.} addition to the LHC event generation tool box. In case
we want to improve the phase space coverage or include subtraction
methods through a pre-defined event weight this is obviously possible.
The same setup will also allow us to generate events from an actual LHC
event sample or to combine actual data with Monte Carlo events for
training, wherever such a thing might come in handy for an analysis or
a fundamental physics question.\bigskip

\begin{center} \textbf{Acknowledgments} \end{center}

We are very grateful to Gregor Kasieczka for his collaboration in the
early phase of the project and to Jonas Glombitza and Till Bungert for
fruitful discussions. We would also like to thank Steffen Schumann for
very helpful physics discussions, asking all the right questions, and
pointing out the similarity of on-shell peaks and phase space
boundaries from a technical point of view. RW acknowledges support by
the IMPRS-PTFS.  The research of AB was supported by the Deutsche
Forschungsgemeinschaft (DFG, German Research Foundation) under grant
396021762 — TRR 257 “Particle Physics Phenomenology after the Higgs
Discovery”.

\end{fmffile}

\bibliography{literature}

\end{document}